
\magnification=1200
\pageno=0 \footline={\ifnum\pageno<1 {\hfil}
\else
              {\hss\tenrm\folio\hss}   \fi}
\hbox to \hsize{\vbox{\hsize 8 em
{\bf \centerline{Groupe d'Annecy}
\ \par
\centerline{Laboratoire}
\centerline{d'Annecy-le-Vieux de}
\centerline{Physique des Particules}}}
\hfill
\hfill
\vbox{\hsize 8 em
{\bf \centerline{Groupe de Lyon}
\ \par
\centerline{   }
\centerline{Ecole Normale}
\centerline{Sup\'erieure de Lyon}}}}
\centerline{   }
\hrule height.42mm

\vskip 20mm

\centerline {{\bf SLAVE-PARTICLE
QUANTIZATION AND SUM RULES IN THE t-
J MODEL}}

\vskip 20mm

\centerline {J.C. LE GUILLOU (*)  and  E.
RAGOUCY}

\bigskip

\centerline { {\it Laboratoire de Physique
Th\'eorique ENSLAPP} (**),}
\centerline { {\it LAPP, B.P. 110, F-74941
Annecy-le-Vieux C\'edex}}

\vfill

\centerline{ {\bf Abstract}}

\bigskip

 In the framework of constrained systems, we
give the classical Hamiltonian formulation
of slave-particle models and their correct
quantization. The electron-momentum distribution
function in the t-J and Hubbard models is then
studied in the framework of slave-particle
approaches and within the decoupling scheme. We
show that criticisms that have been
addressed in this context coming from a violation
of the sum rule for the physical electron are
not valid. Due to the correct quantization rules for
the slave-particles, the sum rule for the
physical electron is indeed obeyed, both exactly
and within the decoupling scheme.

\vskip 20mm

\vfill

\centerline { {\it To be published in Phys. Rev. B
}}

\vskip 20mm

(*) {\it Universit\'e de Savoie} and {\it Institut
Universitaire de France}

(**) URA 1436 du CNRS, associ\'ee  l'ENS
Lyon

\eject

{\bf I - Introduction}

\medskip

\medskip

Several strongly correlated fermionic systems
such as liquid $^3 He$, heavy-fermion
compounds, high-$T_c$ superconductors, and
Kondo systems are the subject of intense
theoretical as well as experimental interest. The
Hubbard model, originally introduced to describe
correlation effects in narrow {\it d}-band
materials, has been put forward as a possible key
to the understanding of high-$T_c$ oxide
superconductivity [1] .

Describing electrons on a lattice with one orbital
per site, the Hubbard Hamiltonian :
$$
\hat H=\sum\limits_{ij,\sigma } {t_{ij}\,\hat
c_{i\sigma }^\dagger \hat c_{j\sigma
}}+\;U\sum\limits_i {\hat n_{i\sigma }\hat n_{i(-
\sigma )}} \eqno (1)
$$
includes a kinetic hopping term $\ t_{ij} \ $
between different sites and an on-site Coulomb
repulsion {\it U}  between electrons of different
spin. The operator $\ \hat c_{i\sigma } \ $
annihilates an electron of spin  $\sigma$  at site
{\it i} and the occupation number operator is $\
\hat n_{i\sigma }=\hat c_{i\sigma }^\dagger \hat
c_{i\sigma }$ .

Of particular interest is the strong-coupling regime
where an effective Hamiltonian of the
Hubbard model with nearest-neighbor hopping is
the t-J model Hamiltonian [2] :
$$
\hat H=t\sum\limits_{<ij>,\sigma } {(\hat
c_{i\sigma }^\dagger \hat c_{j\sigma
}+h.c.)}+\;J\;\sum\limits_{<ij>} {({\hat {\vec
{S_i}}}\cdot {\hat {\vec {S_j}}}-{\textstyle{1
\over 4}}\hat n_i\,\hat n_j)} \eqno (2)
$$
$$
(\;\;{\rm with}\;\;\;J=4t^2/U\;\;\;{\rm and}\;\;\;{\hat
{\vec {S_i}}}={\textstyle{1 \over
2}}\sum\limits_{\alpha \beta } {\hat c_{i\alpha
}^\dagger \vec \sigma _{\alpha \beta }\hat
c_{i\beta }}\;\;)
$$
The on-site Coulomb repulsion there is very large
as compared with the electron-hopping
energy, and therefore when there is less than half
filling the system will avoid configuration
with doubly occupied sites. One has thus the
constraint :
$$
\sum\limits_\sigma  {\hat c_{i\sigma }^\dagger
\hat c_{i\sigma }}\le \;1 \eqno (3)
$$

Apart from numerical approaches, a popular
analytical approach to the t-J model is the
slave-particle theory [3] where for instance in the
slave-fermion representation the electron
operator $\ \hat c_{i\alpha }^\dagger \ $ is written
as $\ \hat c_{i\alpha }^\dagger =\hat b_{i\alpha
}^\dagger \,\hat f_i \ $ , with $\ f \ $ the slave
fermion and $\ b_\alpha \ $ a boson.
Instead of the constraint of Eq. (3) , which is
difficult to handle, one considers the {\it a priori}
more convenient slave-particle constraint avoiding
double occupancy at site {\it i} :
$$
\hat f_i^\dagger \hat f_i+\sum\limits_\alpha  {\hat
b_{i\alpha }^\dagger \hat b_{i\alpha }}=1 \eqno
(4)
$$
With the boson operator $\ \hat b_{i\alpha
}^\dagger \ $ keeping track of the spin and the
fermion operator $\ \hat f_i^\dagger \ $ keeping
track of the charge, this formalism is well adapted
to study the problem of the decoupling of spin and
charge degrees of freedom in the large-{\it U}-
limit Hubbard model. This decoupling,
characteristic of Luttinger liquids, appears to occur
in 1D [4], the situation being still confused in 2D.
At the mean field level, spinons and holons, the
elementary spin and charge excitations, may be
separated in this formalism but are strongly
coupled beyond the mean-field approximation.

Such slave-particle approaches are usually studied
in a functional integral (over coherent
states of Fermi and Bose fields) representation of
the partition function, the slave-particle
constraints being enforced by functional
integration over Lagrange multipliers.

However one may wonder what happens within a
direct operator quantum approach of
such slave-particle theories. Indeed, sum rules,
coming from operator commutation relations,
for spectral functions of the boson and fermion
were used recently in Ref. [5] in a study of the
electron-momentum distribution function in the t-J
model in the framework of the slave-particle
approach and within the decoupling scheme for
the electron Green's function. It was claimed there
that the sum rule for the physical electron was not
obeyed within this framework and
correspondingly that the electron Fermi surface
(EFS) was not explained. However it happens
that the operator commutation relations used in
Ref. [5] for the slave fermion and boson were
the usual naive ones, i.e. the same as if the slave-
particle constraint was not present. One then
can question the use of such operator commutation
relations in a slave-particle approach.

In this paper, we study the direct operator
approach of such slave-particle theories. In
Sec. II, we present at the classical level the
consistent Hamiltonian formulation of models
having a slave-particle constraint for their fields.
We show in Sec. III at the quantum level, for
the slave-fermion and the slave-boson
representations of the t-J model, the modifications
in the
sum rules for the slave particles coming from the
fact that the correct canonical relations
compatible with the constraints are not the naive
ones. We present in Sec. IV a direct explicit
operator quantization of the slave-particle
approaches of the t-J model which confirms the
results obtained in Sec. III . In Sec. V, we extend
our analysis to a slave-boson representation
which has been introduced [6] for the finite-{\it
U} Hubbard model. Sec. VI summarizes our
conclusions, and an Appendix presents
calculations omitted for clarity in Sec. II.

\medskip

\medskip

{\bf II - Hamiltonian formulation for slave
models}

\medskip

\medskip

Let us consider the general classical Lagrangian
(written in real time) for {\it n} bosons $\ b_\alpha
\ $ and {\it m} fermions $\ f_\alpha \ $
(Grassmann variables) on a lattice :
$$
L=i\sum\limits_{i,\alpha } {b_{i\alpha }^\dagger
\partial _tb_{i\alpha }}+i\sum\limits_{i,\sigma }
{f_{i\sigma }^\dagger \partial _tf_{i\sigma
}}+X(b^\dagger ,b,f^\dagger ,f)+\sum\limits_i
{\Lambda _i\;(\;\sum\limits_\alpha  {b_{i\alpha
}^\dagger b_{i\alpha }}+\sum\limits_\sigma
{f_{i\sigma }^\dagger f_{i\sigma }}-1\;)} \eqno
(5)
$$
bosons and fermions being submitted to the slave-
particle constraint at each site {\it i} :
$$
\Phi _i\equiv \sum\limits_\alpha  {b_{i\alpha
}^\dagger b_{i\alpha }}+\sum\limits_\sigma
{f_{i\sigma }^\dagger f_{i\sigma }}-1=0 \eqno
(6)
$$

One shows in the Appendix that, after an {\it \`a la
Dirac} treatment [7], one gets an
Hamiltonian formalism with the non zero brackets
:
$$
\left\{ {b_{i\alpha },b_{j\beta }^\dagger }
\right\}=-\;i\;\delta _{ij}\delta _{\alpha \beta
}\;\;\;\;\;\;\;\;\;\;\left\{ {f_{i\sigma },f_{j\tau
}^\dagger } \right\}=-\;i\;\delta _{ij}\delta
_{\sigma \tau }\;\;\;\;\;\;\;\;\;\;\left\{ {\Lambda _i,\Pi
_j} \right\}=\delta _{ij} \eqno (7)
$$
${\underline {\rm but}}$ with the {\it first class}
[7-9] constraints :
$$
\Phi _i\approx 0\;\;\;\;\;\;\;\;\;\;{\rm
and}\;\;\;\;\;\;\;\;\;\;\Pi _i\approx 0 \eqno (8)
$$
where, following Dirac, first class means here $\
\left\{ {\Phi _i,\Pi _i} \right\}\approx 0 \ $ and
where the symbol $\ \approx 0 \ $ (weakly zero)
means that one has to set the constraints only after
computing all the brackets. We use graded
Poisson-Dirac brackets [9] such that for instance :
$$
\left\{ {B,A} \right\}=-\,(-1)^{ab}\;\left\{ {A,B}
\right\}\;\;\;\;\;\;\;\;\;\;\left\{ {A,BC} \right\}=\left\{
{A,B} \right\}C+\,(-1)^{ab}\;B\left\{ {A,C}
\right\} \eqno (9)
$$
where {\it a} = 0  if  {\it A}  is a bosonic quantity
and {\it a} = 1  if  {\it A}  is a fermionic quantity.

The role of the first class constraints is to generate
infinitesimal contact transformations
(that we shall call as usual [8] gauge
transformations) in the Hamiltonian formalism that
do not
affect the physical state of the system. We have
here for the non-zero brackets of the
fundamental variables with the first class
constraints :
$$
\eqalignno{&\left\{ {b_{i\alpha },\Phi _j}
\right\}=-\;i\;b_{i\alpha }\delta
_{ij}\;\;\;\;\;\;\;\;\;\;\;\;\left\{ {b_{i\alpha }^\dagger
,\Phi _j} \right\}=+\;i\;b_{i\alpha }^\dagger \delta
_{ij}&(10)\cr
  &\left\{ {f_{i\sigma },\Phi _j} \right\}=-
\;i\;f_{i\sigma }\delta _{ij}\;\;\;\;\;\;\;\;\;\;\;\;\left\{
{f_{i\sigma }^\dagger ,\Phi _j}
\right\}=+\;i\;f_{i\sigma }^\dagger \delta
_{ij}&(11)\cr
  &\left\{ {\Lambda _i,\Pi _j} \right\}=\delta
_{ij}&(12)\cr}
$$
which means that the gauge transformations are
$\;b\to e^{-i\theta }b\;$,$\;b^\dagger \to e^{i\theta
}b^\dagger \;$,$\;f\to e^{-i\theta }f\;$,$\;f^\dagger
\to e^{i\theta }f^\dagger \;$,$\;\Lambda \to
\Lambda +a\;$,$\;\Pi \to \Pi \;$ .

The standard strategy [8] is then to fix the gauge
by choosing explicit forms for each
gauge and imposing them as constraints not
following from the Lagrangian. The choice of
gauges should be made in such a way that the
constraints $\ \Phi _i \ $ and $\ \Pi _i \ $ will cease
to be first class. It happens that a convenient
choice here is the linear one :
$$
\Phi '_i\equiv \sum\limits_\alpha  {(\;G_{i\alpha
}b_{i\alpha }+\;G_{i\alpha }^\dagger b_{i\alpha
}^\dagger \;)}+\sum\limits_\sigma  {(\;H_{i\sigma
}f_{i\sigma }-\;H_{i\sigma }^\dagger f_{i\sigma
}^\dagger \;)}+K_i\;\;\approx \;\;0 \eqno (13)
$$
$$
\Lambda _i\;\;\approx \;\;0 \eqno (14)
$$
where the {\it G} 's, the {\it K} 's and the
(Grassmannian) {\it H} 's are parameters.

We have then for the non-zero brackets of the
fundamental variables with these gauge
fixing constraints :
$$
\eqalignno{&\left\{ {b_{i\alpha },\Phi '_j}
\right\}=-\;i\;G_{i\alpha }^\dagger \delta
_{ij}\;\;\;\;\;\;\;\;\;\;\;\;\left\{ {b_{i\alpha }^\dagger
,\Phi '_j} \right\}=+\;i\;G_{i\alpha }\delta
_{ij}&(15)\cr
  &\left\{ {f_{i\sigma },\Phi '_j} \right\}=-
\;i\;H_{i\sigma }^\dagger \delta
_{ij}\;\;\;\;\;\;\;\;\;\;\;\;\left\{ {f_{i\sigma }^\dagger
,\Phi '_j} \right\}=+\;i\;H_{i\sigma }\delta
_{ij}&(16)\cr
  &\left\{ {\Pi _i,\Lambda _j} \right\}=-\;\delta
_{ij}&(17)\cr}
$$
and we obtain the bracket of the slave-particle
constraint with its gauge fixing constraint as :
$$
\left\{ {\Phi _i,\Phi '_j}
\right\}=i\;\left[\;\sum\limits_\alpha  {(\;G_{i\alpha
}b_{i\alpha }-\;G_{i\alpha }^\dagger b_{i\alpha
}^\dagger \;)}+\sum\limits_\sigma  {(\;H_{i\sigma
}f_{i\sigma }+\;H_{i\sigma }^\dagger f_{i\sigma
}^\dagger \;)}\;\right]\;\delta _{ij}\equiv
i\;D_i\;\delta _{ij} \eqno (18)
$$
which is not weakly zero.

Let us note that the Hamiltonian (see the
Appendix) is now :
$$
H_3=-\,X(b^\dagger ,b,f^\dagger ,f)-
\sum\limits_i {(x_i-\Lambda
_i)\;(\;\sum\limits_\alpha  {b_{i\alpha }^\dagger
b_{i\alpha }}+\sum\limits_\sigma  {f_{i\sigma
}^\dagger f_{i\sigma }}-1\;)}+\sum\limits_i {\Pi
_iw_i} \eqno (19)
$$
where the first class constraints coefficients  {\it
x}  and  {\it w}  (which are in $H_3$ independent
of the fields) are determined by the requirement
that the time derivative of the gauge fixing
constraints is weakly zero :
$$
\left\{ {\Phi '_i,H_3} \right\}=i\sum\limits_\alpha
{(G_{i\alpha }{{\partial X} \over {\partial
b_{i\alpha }^\dagger }}-\;G_{i\alpha }^\dagger
{{\partial X} \over {\partial b_{i\alpha
}}})}\;+i\sum\limits_\sigma  {(H_{i\sigma
}{{\partial X} \over {\partial f_{i\sigma }^\dagger
}}-\;H_{i\sigma }^\dagger {{\partial X} \over
{\partial f_{i\sigma }}})}-\;i\,(x_i-\Lambda
_i)\;D_i \approx 0 \eqno (20)
$$
$$\left\{ {\Lambda _i,H_3} \right\}=w_i \approx
0 \eqno (21)$$

Defining $\ \varphi _1\equiv \Phi _i\;\;,\;\;\varphi
_2\equiv \Phi '_i\;\;,\;\;\varphi _3\equiv \Pi
_i\;\;,\;\;\varphi _4\equiv \Lambda _i \ $ , the
matrix $\ C_{ab}\equiv \left\{ {\varphi _a,\varphi
_b} \right\} \ $ is non singular and all the
constraints are now {\it second class} [7-9].
Systematic use of the standard Dirac bracket [7] of
two quantities  {\it A}  and  {\it B}  :
$$
\left\{ {A,B} \right\}_*\equiv \left\{ {A,B}
\right\}-\sum\limits_{a,b} {\left\{ {A,\varphi _a}
\right\}\,(C^{-1})_{ab}\,\left\{ {\varphi _b,B}
\right\}} \eqno (22)
$$
then allows one to set all these second class
constraints strongly to zero because the Dirac
bracket of anything with a second class constraint
vanishes.

We thus obtain the correct classical non zero
canonical relations for slave-particle models,
compatible with the constraints :
$$\eqalignno{&i\;\left\{ {b_{i\alpha },b_{j\beta
}^\dagger } \right\}_*=\;\;\;\,[\;\delta _{\alpha
\beta }-\;(\;G_{i\beta }b_{i\alpha }-\;G_{i\alpha
}^\dagger b_{i\beta }^\dagger \;)/D_i\;]\;\delta
_{ij}&(23)\cr
  &i\;\left\{ {b_{i\alpha }^\dagger ,b_{j\beta
}^\dagger } \right\}_*=\;\;\;\,[\;(\;G_{i\beta
}b_{i\alpha }^\dagger -\;G_{i\alpha
}^{}b_{i\beta }^\dagger \;)/D_i\;]\;\delta
_{ij}&(24)\cr
  &i\;\left\{ {b_{i\alpha }^{},b_{j\beta }^{}}
\right\}_*=\;\;\;\,[\;(\;G_{i\beta }^\dagger
b_{i\alpha }^{}-\;G_{i\alpha }^\dagger b_{i\beta
}^{}\;)/D_i\;]\;\delta _{ij}&(25)\cr
  &i\;\left\{ {f_{i\sigma },f_{j\tau }^\dagger }
\right\}_*=\;\;\;\,[\;\delta _{\sigma \tau
}+\;(\;H_{i\tau }f_{i\sigma }+\;H_{i\sigma
}^\dagger f_{i\tau }^\dagger \;)/D_i\;]\;\delta
_{ij}&(26)\cr
  &i\;\left\{ {f_{i\sigma }^\dagger ,f_{j\tau
}^\dagger } \right\}_*=-\;[\;(\;H_{i\tau
}f_{i\sigma }^\dagger +\;H_{i\sigma
}^{}f_{i\tau }^\dagger \;)/D_i\;]\;\delta
_{ij}&(27)\cr
  &i\;\left\{ {f_{i\sigma }^{},f_{j\tau }^{}}
\right\}_*=-\;[\;(\;H_{i\tau }^\dagger f_{i\sigma
}^{}+\;H_{i\sigma }^\dagger f_{i\tau
}^{}\;)/D_i\;]\;\delta _{ij}&(28)\cr
  &i\;\left\{ {f_{i\sigma }^{},b_{j\alpha
}^\dagger } \right\}_*=-\;[\;(\;G_{i\alpha
}^{}f_{i\sigma }^{}-\;H_{i\sigma }^\dagger
b_{i\alpha }^\dagger \;)/D_i\;]\;\delta
_{ij}&(29)\cr
  &i\;\left\{ {f_{i\sigma }^\dagger ,b_{j\alpha
}^{}} \right\}_*=-\;[\;(\;G_{i\alpha }^\dagger
f_{i\sigma }^\dagger -\;H_{i\sigma
}^{}b_{i\alpha }^{}\;)/D_i\;]\;\delta _{ij}&(30)\cr
  &i\;\left\{ {f_{i\sigma }^\dagger ,b_{j\alpha
}^\dagger } \right\}_*=\;\;\;\,[\;(\;G_{i\alpha
}^{}f_{i\sigma }^\dagger -\;H_{i\sigma
}^{}b_{i\alpha }^\dagger \;)/D_i\;]\;\delta
_{ij}&(31)\cr
  &i\;\left\{ {f_{i\sigma }^{},b_{j\alpha }^{}}
\right\}_*=\;\;\;\,[\;(\;G_{i\alpha }^\dagger
f_{i\sigma }^{}-\;H_{i\sigma }^\dagger
b_{i\alpha }^{}\;)/D_i\;]\;\delta _{ij}&(32)\cr}$$
It is then clear, as we shall explicitly show below,
that a correct operator quantization of such
slave-particle models must be based on these new
canonical relations, and not on the naive
ones of Eq. (7) which are not compatible with the
constraints.

\medskip

Let us note that we have $\ \left\{ {\Pi _i,\Lambda
_i} \right\}_*=\left\{ {\Pi _i,\Pi _i}
\right\}_*=\left\{ {\Lambda _i,\Lambda _i}
\right\}_*=0 \ $ indeed compatible with
the constraints $\ \Pi _i=\Lambda _i=0 \ $. On the
other hand the Hamiltonian is finally :
$$
H_4=-\,X(b^\dagger ,b,f^\dagger ,f) \eqno (33)
$$
and we have consistently for example $\ \left\{
{b_{i\alpha },H_3} \right\}=\left\{ {b_{i\alpha
},H_4} \right\}_* \ $ since :
$$
\eqalignno{&\left\{ {b_{i\alpha },H_3}
\right\}=i{{\partial X} \over {\partial b_{i\alpha
}^\dagger }}-\;i\;(x_i-\Lambda _i)b_{i\alpha
}&(34)\cr
  &\left\{ {b_{i\alpha },H_4}
\right\}_*=i{{\partial X} \over {\partial b_{i\alpha
}^\dagger }}-i{{b_{i\alpha }} \over {D_i}}\left[
{\sum\limits_\beta  {(G_{i\beta }{{\partial X}
\over {\partial b_{i\beta }^\dagger }}-\;G_{i\beta
}^\dagger {{\partial X} \over {\partial b_{i\beta
}}})}+\sum\limits_\sigma  {(H_{i\sigma
}{{\partial X} \over {\partial f_{i\sigma }^\dagger
}}-H_{i\sigma }^\dagger {{\partial X} \over
{\partial f_{i\sigma }}})}} \right]&(35)\cr}
$$
which are indeed equal using Eq. (20).

\medskip

One may wonder at this stage about what happens
for the classical version of the $\ Sl(1\,|2) \ $
superalgebra obeyed in the t-J model [10] by the
Hubbard [11] operators $\ X_i^{ab} \ $ :
$$
\left[ {X_i^{ab},X_j^{cd}} \right]_\pm
=(X_i^{ad}\delta _{bc}\pm X_i^{cb}\delta
_{ad})\;\delta _{ij}\;\;\;\;\;\;\;\;{\rm
with}\;\;\;X_i^{00}+\sum\limits_\alpha
{X_i^{\alpha \alpha }}=1 \eqno (36)
$$
which opens the way for a supersymmetric t-J
model [10] and which is verified in the literature
within a slave representation using the naive
canonical relations. In fact, taking for instance in
the t-J model the slave-fermion representation $\
c_{i\alpha }^\dagger =b_{i\alpha }^\dagger \,f_i \
$ of the electron field $\ c_{i\alpha }^\dagger \ $
which corresponds to $\ X_i^{\alpha 0} \ $, the $\
Sl(1\,|2) \ $ superalgebra is obeyed using either
the naive initial brackets of Eq. (7) or the Dirac
brackets of Eqs. (23-32). For example one has :
$$
i\;\left\{ {c_{i\alpha }^\dagger ,c_{j\beta }^{}}
\right\}=i\;\left\{ {c_{i\alpha }^\dagger ,c_{j\beta
}^{}} \right\}_*=\;\;[\;b_{i\alpha }^\dagger
b_{i\beta }+\;f_i^\dagger f_i\;\delta _{\alpha \beta
}\;]\;\delta _{ij} \eqno (37)
$$
where $\ b_{i\alpha }^\dagger b_{i\beta }\ $
corresponds to $\ X_i^{\alpha \beta }\ $ and
 $\ f_i^\dagger f_i\ $ to $\ X_i^{00}$. The reason
for this property is that the Dirac bracket of any
two gauge invariant quantities is the same as their
initial bracket. As all the
generators of the $\ Sl(1\,|2) \ $ superalgebra are
gauge invariant, it is licit to use the initial brackets
of the {\it b}'s and {\it f}'s to compute the Dirac
brackets of these generators. One must however
realize that this property is valid only for gauge
invariant quantities. At the quantum level, a proper
quantization should inherit the same property for
(anti)commutators of gauge invariant quantities,
but not for products of these quantities. Let us
also note that, using the slave-particle constraint,
one has from Eq.(37):
$$
\sum\limits_\alpha  {i\;\left\{ {c_{i\alpha
}^\dagger ,c_{j\alpha }^{}}
\right\}}=\sum\limits_\alpha  {i\;\left\{ {c_{i\alpha
}^\dagger ,c_{j\alpha }^{}}
\right\}_*}=\;\;[\;1+\;f_i^\dagger f_i\;]\;\delta _{ij}
\eqno (38)
$$
a relation that we shall recover below in a sum rule
at the quantum level.

\medskip

\medskip

{\bf III - Modifications of the naive sum rules for
the slave t-J models}

\medskip

\medskip

As mentioned in the Introduction, it has been
stressed in the literature [5] that a study of
the electron-momentum distribution function in the
t-J model in the framework of the slave-particle
approach and within the decoupling scheme would
give rise to a violation of the sum
rule of electron number. However, to obtain this
result, sum rules for the slave particles using
quantization of the naive relations of Eq. (7) were
used. On the contrary we shall show in this
Section that starting from the correct canonical
relations compatible with the constraints
produces modifications in these sum rules for the
slave particles in such a way that the sum rule
of electron number is indeed obeyed. Though we
shall present in the next Section a direct
explicit quantization which confirms this result,
we think that it is important for the clarity of the
exposition to see here the simple structure of these
modifications in the sum rules.

\medskip

{\bf a) The slave-fermion representation}

\medskip

Let us consider in the t-J model the slave-fermion
representation $\ \hat c_{i\alpha }^\dagger =\hat
b_{i\alpha }^\dagger \,\hat f_i \ $ of the electron
operator $\ \hat c_{i\alpha }^\dagger \ $ , with the
slave-particle constraint avoiding double
occupancy at site {\it i} :
$$
\hat f_i^\dagger \hat f_i+\sum\limits_\alpha  {\hat
b_{i\alpha }^\dagger \hat b_{i\alpha }}=1 \eqno
(39)
$$
As in Ref. [5], we assume that there is no Bose
condensation of spinons i.e. the temperature of
the system is $0^+$ . The hole-doping
concentration $\ \delta \ $ is given by $\ \delta
=\left\langle {\hat f_i^\dagger \hat f_i}
\right\rangle \ $ .

For discussing the electron-momentum
distribution function, the Matsubara electron
Green's function in imaginary time [12,13] :
$$
E_\alpha (r,\tau )=-\;\left\langle {\,T_\tau \,(\,\hat
f_i^\dagger (\tau )\,\hat b_{i\alpha }(\tau )\,\hat
b_{j\alpha }^\dagger (0)\,\hat f_j(0)\,)\,}
\right\rangle\;\;\;\;\;\;\;\;\;\;\;\;(r=i-j)\;\;, \eqno (40)
$$
where $\ \left\langle {\,...\,} \right\rangle \ $
means the thermodynamical average, was
considered in Ref. [5] within the decoupling
approximation written on Fourier transform :
$$
E_\alpha (k,\omega _n)={\textstyle{1 \over
N}}\sum\limits_q {{\textstyle{1 \over \beta
}}}\sum\limits_{\omega _m} {F(q,\omega
_m)\;B_\alpha (q+k,\omega _m+\omega _n)}
\eqno (41)
$$
{\it F}  is the Green's function for the slave-
fermion  {\it f}  and $\ B_\alpha \ $ is the Green's
function for the boson $\ b_\alpha \ $ .
Introducing the Lehmann's spectral
representations :
$$
\left( {\matrix{{E_\alpha }\cr
F\cr
{B_\alpha }\cr
}} \right)(k,\omega _n)=\int_{-\infty }^{+\infty }
{{\textstyle{{d\omega } \over {2\pi }}}\,\left(
{\matrix{{A_{e\alpha }}\cr
{A_f}\cr
{A_{b\alpha }}\cr
}} \right)(k,\omega )\,{1 \over {(i\omega _n-
\omega )}}} \eqno (42)
$$
one easily obtains the electron spectral function as
:
$$
A_{e\alpha }(k,\omega )={\textstyle{1 \over
N}}\sum\limits_q {\int_{-\infty }^{+\infty }
{{\textstyle{{d\omega '} \over {2\pi
}}}\,A_f(q,\omega ')\,A_{b\alpha }(q+k,\omega
+\omega ')\,[\,n_F(\omega ')+n_B(\omega
+\omega ')\,]}} \eqno (43)
$$
where $\ n_F \ $ and $\ n_B \ $ are respectively
the Fermi and Bose distribution functions.
Using the expressions for the numbers of slave-
fermions, bosons and electrons in state {\it q} :
$$
\eqalignno{&n_f(q)=\int_{-\infty }^{+\infty }
{{\textstyle{{d\omega } \over {2\pi
}}}\,n_F(\omega )\,A_f(q,\omega
)}\;\;\;\;\;\;\;\;\;\;n_{b\alpha }(q)=\int_{-\infty
}^{+\infty } {{\textstyle{{d\omega } \over {2\pi
}}}\,n_B(\omega )\,A_{b\alpha }(q,\omega )}\cr
  &n_e(q)=\sum\limits_\alpha  {\int_{-\infty
}^{+\infty } {{\textstyle{{d\omega } \over {2\pi
}}}\,n_F(\omega )\,A_{e\alpha }(q,\omega
)}}\;\;\;\;\;,&(44)\cr}
$$
the identity : $\ n_F(\omega )\;[\,n_F(\omega
')+n_B(\omega +\omega ')\,]=n_B(\omega
+\omega ')\;[\,1-n_F(\omega ')\,] \ $ , and the
sum rules :
$$
\int_{-\infty }^{+\infty } {{\textstyle{{d\omega }
\over {2\pi }}}\,A_f(q,\omega )}=\left\langle
{\;\left[ {\hat f_q,\hat f_q^\dagger } \right]_+}
\right\rangle\;\;\;\;\;\;\;\sum\limits_\alpha  {\int_{-
\infty }^{+\infty } {{\textstyle{{d\omega } \over
{2\pi }}}\,A_{b\alpha }(q,\omega )}}=\left\langle
{\,\sum\limits_\alpha  {\,\left[ {\hat b_{q\alpha
},\hat b_{q\alpha }^\dagger } \right]_-}}
\right\rangle \eqno (45)
$$
one obtains the sum rule for the electron spectral
function :
$$
\eqalignno{&\sum\limits_\alpha  {\int_{-\infty
}^{+\infty } {{\textstyle{{d\omega } \over {2\pi
}}}\,A_{e\alpha }(k,\omega )}}\cr
  &={\textstyle{1 \over N}}\sum\limits_q
{n_f(q)\;\left\langle {\,\sum\limits_\alpha  {\,\left[
{\hat b_{(q+k)\alpha },\hat b_{(q+k)\alpha
}^\dagger } \right]_-}}
\right\rangle}+{\textstyle{1 \over
N}}\sum\limits_{q,\alpha } {n_{b\alpha
}(q+k)\;\left\langle {\;\left[ {\hat f_q,\hat
f_q^\dagger } \right]_+} \right\rangle}&(46)\cr}
$$
and the expression for the number of electrons in
state  {\it k}  :
$$
n_e(k)={\textstyle{1 \over
N}}\sum\limits_{q,\alpha } {n_{b\alpha
}(q+k)\;\left\langle {\;\left[ {\hat f_q,\hat
f_q^\dagger } \right]_+} \right\rangle}\;\;-
\;\;{\textstyle{1 \over N}}\sum\limits_{q,\alpha }
{n_{b\alpha }(q+k)\;n_f(q)} \eqno (47)
$$

Let us make the remark, not mentioned in Ref. [5]
, that the last equation can in fact be
directly derived from the very definition of the
decoupling approximation :
$$
E_\alpha (r,\tau )=B_\alpha (r,\tau )\;F(-r,-\tau )
\eqno (48)
$$
where we used :
$$
\left\langle {\,T_\tau \,(\,\hat b_{i\alpha }(\tau
)\,\hat b_{j\alpha }^\dagger (0)\,)\,}
\right\rangle=-\;B_\alpha (r,\tau
)\;\;\;\;\;\;\;\;\;\;\left\langle {\,T_\tau \,(\,\hat
f_i^\dagger (\tau )\,\hat f_j(0)\,)\,}
\right\rangle=F(-r,-\tau ) \eqno (49)
$$
Then, with $\ \eta \to 0^+ \ $ [12] , one has :
$$
n_e={\textstyle{1 \over N}}\sum\limits_{i,\alpha
} {E_\alpha (0,-\eta )}={\textstyle{1 \over
N}}\sum\limits_{i,\alpha } {B_\alpha (0,-\eta
)\;F(0,\eta )}={\textstyle{1 \over N}}\sum\limits_i
{(\sum\limits_\alpha  {\left\langle {\,\hat
b_{i\alpha }^\dagger \hat b_{i\alpha }}
\right\rangle\,})\;\left\langle {\,\hat f_i\hat
f_i^\dagger } \right\rangle} \eqno (50)
$$
using $\ B_\alpha (0,-\eta )=-\left\langle {\,\hat
b_{i\alpha }^\dagger \hat b_{i\alpha }}
\right\rangle \ $ and $\ F(0,\eta )=-\left\langle
{\,\hat f_i\hat f_i^\dagger } \right\rangle \ $. This
gives in Fourier transform:
$$
n_e(k)={\textstyle{1 \over
N}}\sum\limits_{q,\alpha } {B_{(q+k)\alpha }(-
\eta )\;F_q(\eta )}={\textstyle{1 \over
N}}\sum\limits_{q,\alpha } {\left\langle {\,\hat
b_{(q+k)\alpha }^\dagger \hat b_{(q+k)\alpha }}
\right\rangle\;\left\langle {\,\hat f_q\hat
f_q^\dagger } \right\rangle} \eqno (51)
$$
expression which, using the anticommutator of the
fermion, can immediately be written in the
form of Eq. (47) .

The problem now is to evaluate the
thermodynamical average of the commutator and
anticommutator entering in these expressions.
Using the quantization from the Dirac brackets :
$$
i\left\{ {A,B} \right\}_*\;\;\to \;\;\;\left[ {\hat A,\hat
B} \right]_\pm  \eqno (52)
$$
one obtains from our results Eqs. (23) and (26) of
the preceding Section the form of the
commutator and anticommutator at equal times :
$$
\left[ {\hat b_{i\alpha },\hat b_{j\alpha }^\dagger
} \right]_-=(\;1-\;\hat \Theta _{i\alpha }\;)\;\delta
_{ij}\;\;\;\;\;\;\;\;\;\;\;\;\;\;\;\;\;\;\;\left[ {\hat f_i,\hat
f_j^\dagger } \right]_+=(\;1+\;\hat \Theta
_i\;)\;\delta _{ij} \eqno (53)
$$
where we have introduced the $\ \hat \Theta \ $
operators defined through the quantizations :
$$
(\;G_{i\alpha }b_{i\alpha }-\;G_{i\alpha
}^\dagger b_{i\alpha }^\dagger \;)/D_i\;\;\equiv
\Theta _{i\alpha }\;\;\to \;\;\hat \Theta _{i\alpha
}\;\;\;\;\;\;\;\;\;\;\,(\;H_if_i+\;H_i^\dagger
f_i^\dagger \;)/D_i\;\;\equiv \Theta _i\;\;\to \;\;\hat
\Theta _i \eqno (54)
$$
Let us note that the definition of $\ D_i \ $ from
Eq. (18) gives by quantization the relation :
$$
\sum\limits_\alpha  {\hat \Theta _{i\alpha }}+\hat
\Theta _i=1 \eqno (55)
$$
With $\ \hat \Theta _n=\sum\nolimits_p {e^{\kern
1pti\kern 1ptp\kern 1ptn}\,\hat \Xi _p} \ $
 and $\ \hat \Theta _{n\alpha }=\sum\nolimits_p
{e^{\kern 1pti\kern 1ptp\kern 1ptn}\,\hat \Xi
_{p\alpha }} \ $ , it follows that :
$$
\eqalignno{&\left\langle {\;\left[ {\hat f_q,\hat
f_q^\dagger } \right]_+}
\right\rangle=1+\;\left\langle {\,\hat \Xi _0}
\right\rangle &(56)\cr
  &\left\langle {\,\sum\limits_\alpha  {\,\left[ {\hat
b_{(q+k)\alpha },\hat b_{(q+k)\alpha }^\dagger }
\right]_-}} \right\rangle=2-\;\left\langle
{\,\sum\limits_\alpha  {\,\hat \Xi _{0\alpha }}}
\right\rangle=1+\;\left\langle {\,\hat \Xi _0}
\right\rangle &(57)\cr}
$$

Since with the hole-doping concentration $\ \delta \
$ one has :
$$
{\textstyle{1 \over N}}\sum\limits_q
{n_f(q)}=\delta \;\;\;\;\;\;\;\;\;\;\;\;\;{\textstyle{1
\over N}}\sum\limits_{q,\alpha } {n_{b\alpha
}(q)}=1-\delta  \eqno (58)
$$
one obtains the following results :
$$
\eqalignno{&\sum\limits_\alpha  {\int_{-\infty
}^{+\infty } {{\textstyle{{d\omega } \over {2\pi
}}}\,A_{e\alpha }(k,\omega )}}=1+\;\left\langle
{\,\hat \Xi _0} \right\rangle &(59)\cr
  &n_e(k)=(1-\delta )\;(\;1+\;\left\langle {\,\hat \Xi
_0} \right\rangle\;)\;\;-\;\;{\textstyle{1 \over
N}}\sum\limits_{q,\alpha } {n_{b\alpha
}(q+k)\;n_f(q)}&(60)\cr
  &{\textstyle{1 \over N}}\sum\limits_k
{n_e(k)}=(1-\delta )\;(\;1+\;\left\langle {\,\hat \Xi
_0} \right\rangle\;-\delta \;)&(61)\cr}
$$

One then sees clearly the modifications of the sum
rules and of the results of Ref. [5]
coming from the presence of the $\ \hat \Theta \ $
's in the correct canonical relations. The main
point is that our results show that one indeed
obtains the expected result that, if $\ \delta \ $
holes are introduced into the half-filled system, the
total electron number per site would be
$\ (1-\delta ) \ $, instead of the $\ (1-\delta )^2 \ $
found in Ref. [5] .

In fact, from the quantization of Eq. (38) at the
end of the preceding Section, the sum rule
for the electron spectral function must be :
$$
\sum\limits_\alpha  {\int_{-\infty }^{+\infty }
{{\textstyle{{d\omega } \over {2\pi
}}}\,A_{e\alpha }(q,\omega )}}=\left\langle
{\,\sum\limits_\alpha  {\,\left[ {\hat c_{q\alpha
},\hat c_{q\alpha }^\dagger } \right]_+}}
\right\rangle=1+{\textstyle{1 \over
N}}\sum\limits_k {\left\langle {\hat f_k^\dagger
\hat f_k} \right\rangle}=1+\delta  \eqno (62)
$$
being unchanged with respect to Ref. [5].
Comparing with Eq. (59), this gives :
$$
\left\langle {\,\hat \Xi _0} \right\rangle=\delta
\eqno (63)
$$
 From our results, the expected sum rule of the
electron number is indeed recovered with
$\ \left\langle {\,\hat \Xi _0} \right\rangle=\delta \
$ , since $\ {\textstyle{1 \over
N}}\sum\nolimits_k {n_e(k)}=(1-\delta
)\;(\;1+\;\left\langle {\,\hat \Xi _0} \right\rangle\;-
\delta \;)=(1-\delta ) \ $ . However this does not
guarantee the existence of an EFS within the
decoupling scheme in the slave-fermion approach
of the t-J model, and the arguments of Ref. [5]
against an EFS still apply.

\medskip

The electron operator anticommutator furthermore
shows that one has not only the
average equality $\ \left\langle {\,\hat \Theta _i}
\right\rangle=\left\langle {\,\hat \Xi _0}
\right\rangle=\delta \ $ but also the operator
equality $\ \hat \Theta _i=\hat f_i^\dagger \hat f_i \
$
 :  on one hand quantization of Eq. (38) gives $\
\sum\nolimits_\alpha  {\left[ {\hat c_{i\alpha },\hat
c_{i\alpha }^\dagger } \right]_+}=1+\hat
f_i^\dagger \hat f_i \ $ , while on the other hand
one gets by expressing each term of the
anticommutator with the boson and fermion
operators
$$
\sum\nolimits_\alpha  {\left[ {\hat c_{i\alpha },\hat
c_{i\alpha }^\dagger } \right]_+}=1+\hat
f_i^\dagger \hat f_i+\sum\nolimits_\alpha  {\hat
b_{i\alpha }^\dagger (\hat \Theta _i-\hat
f_i^\dagger \hat f_i)\hat b_{i\alpha }}+\hat
f_i^\dagger (\hat \Theta _i-\hat f_i^\dagger \hat
f_i)\hat f_i \eqno (64)
$$
effectively leading to a consistent quantization
expressed by $\ \hat \Theta _i=\hat f_i^\dagger \hat
f_i \ $ .

\medskip

{\bf b) The slave-boson representation}

\medskip

Our analysis of the t-J model in the slave-boson
representation $\ \hat c_{i\sigma }^\dagger =\hat
f_{i\sigma }^\dagger \,\hat b_i \ $ of the electron
operator $\ \hat c_{i\sigma }^\dagger  \ $ , with
the slave-particle constraint avoiding double
occupancy at site  {\it i}  :
$$
\hat b_i^\dagger \hat b_i+\sum\limits_\sigma  {\hat
f_{i\sigma }^\dagger \hat f_{i\sigma }}=1 \eqno
(65)
$$
follows along the same lines. As in Ref. [5], we
assume that there is no Bose condensation of
holons. The hole-doping concentration $\ \delta  \
$ is given by $\ \delta =\left\langle {\hat
b_i^\dagger \hat b_i} \right\rangle \ $ .

The Matsubara electron Green's function in
imaginary time :
$$
E_\sigma (r,\tau )=-\;\left\langle {\,T_\tau \,(\,\hat
b_i^\dagger (\tau )\,\hat f_{i\sigma }(\tau )\,\hat
f_{j\sigma }^\dagger (0)\,\hat b_j(0)\,)\,}
\right\rangle \eqno (66)
$$
within the decoupling approximation written on
Fourier transform reads :
$$
E_\sigma (k,\omega _n)=-\,{\textstyle{1 \over
N}}\sum\limits_q {{\textstyle{1 \over \beta
}}}\sum\limits_{\omega _m} {B(q,\omega _m-
\omega _n)\;F_\sigma (q+k,\omega _m)} \eqno
(67)
$$
and the electron spectral function is :
$$
A_{e\sigma }(k,\omega )={\textstyle{1 \over
N}}\sum\limits_q {\int_{-\infty }^{+\infty }
{{\textstyle{{d\omega '} \over {2\pi
}}}\,A_b(q,\omega '-\omega )\,A_{f\sigma
}(q+k,\omega ')\,[\,n_F(\omega ')+n_B(\omega
'-\omega )\,]}} \eqno (68)
$$
Using the identity $\ n_F(\omega )\;[\,n_F(\omega
')+n_B(\omega '-\omega )\,]=n_F(\omega
')\;[\,1+n_B(\omega '-\omega )\,] \ $ and the sum
rules analogous to Eq. (45) , one obtains the sum
rule for the electron spectral function :
$$
\eqalignno{&\sum\limits_\sigma  {\int_{-\infty
}^{+\infty } {{\textstyle{{d\omega } \over {2\pi
}}}\,A_{e\sigma }(k,\omega )}}\cr
  &={\textstyle{1 \over N}}\sum\limits_q
{n_b(q)\;\left\langle {\,\sum\limits_\sigma  {\,\left[
{\hat f_{(q+k)\sigma },\hat f_{(q+k)\sigma
}^\dagger } \right]_+}}
\right\rangle}+{\textstyle{1 \over
N}}\sum\limits_{q,\sigma } {n_{f\sigma
}(q+k)\;\left\langle {\;\left[ {\hat b_q,\hat
b_q^\dagger } \right]_-} \right\rangle} &(69)\cr}
$$
and the expression for the number of electrons in
state  {\it k}  :
$$
n_e(k)={\textstyle{1 \over
N}}\sum\limits_{q,\sigma } {n_{f\sigma
}(q+k)\;\left\langle {\;\left[ {\hat b_q,\hat
b_q^\dagger } \right]_-}
\right\rangle}\;\;+\;\;{\textstyle{1 \over
N}}\sum\limits_{q,\sigma } {n_{f\sigma
}(q+k)\;n_b(q)} \eqno (70)
$$

Again, the last equation can in fact be directly
derived from the very definition of the
decoupling approximation :
$$
E_\sigma (r,\tau )=-\;F_\sigma (r,\tau )\;B(-r,-\tau
) \eqno (71)
$$
where we used :
$$
\left\langle {\,T_\tau \,(\,\hat f_{i\sigma }(\tau
)\,\hat f_{j\sigma }^\dagger (0)\,)\,}
\right\rangle=-\;F_\sigma (r,\tau
)\;\;\;\;\;\;\;\;\left\langle {\,T_\tau \,(\,\hat
b_i^\dagger (\tau )\,\hat b_j(0)\,)\,} \right\rangle=-
\;B(-r,-\tau ) \eqno (72)
$$
Then, with $\ \eta \to 0^+ \ $ , one has :
$$
n_e={\textstyle{1 \over N}}\sum\limits_{i,\sigma
} {E_\sigma (0,-\eta )}=-\;{\textstyle{1 \over
N}}\sum\limits_{i,\sigma } {F_\sigma (0,-\eta
)\;B(0,\eta )}={\textstyle{1 \over
N}}\sum\limits_i {(\sum\limits_\sigma
{\left\langle {\,\hat f_{i\sigma }^\dagger \hat
f_{i\sigma }} \right\rangle\,})\;\left\langle {\,\hat
b_i\hat b_i^\dagger } \right\rangle} \eqno (73)
$$
using $\ F_\alpha (0,-\eta )=+\;\left\langle {\,\hat
f_{i\sigma }^\dagger \hat f_{i\sigma }}
\right\rangle \ $ and $\ B(0,\eta )=-\;\left\langle
{\,\hat b_i\hat b_i^\dagger } \right\rangle \ $
 . This gives in Fourier transform :
$$
n_e(k)=-\;{\textstyle{1 \over
N}}\sum\limits_{q,\sigma } {F_{(q+k)\sigma }(-
\eta )\;B_q(\eta )}={\textstyle{1 \over
N}}\sum\limits_{q,\sigma } {\left\langle {\,\hat
f_{(q+k)\sigma }^\dagger \hat f_{(q+k)\sigma }}
\right\rangle\;\left\langle {\,\hat b_q\hat
b_q^\dagger } \right\rangle} \eqno (74)
$$
expression which, using the commutator of the
boson, can immediately be written in the form
of Eq. (70) .

\medskip

Now, with the quantizations :
$$
(\;G_ib_i-\;G_i^\dagger b_i^\dagger \;)/D_i\;\;\to
\;\;\hat \Theta _i\;\;\;\;\;\;(\;H_{i\sigma }f_{i\sigma
}+\;H_{i\sigma }^\dagger f_{i\sigma }^\dagger
\;)/D_i\;\;\to \;\;\hat \Theta _{i\sigma
}\;\;\;\;\;\;\sum\limits_\sigma  {\hat \Theta
_{i\sigma }}+\hat \Theta _i=1 \eqno (75)
$$
one obtains from our results of the preceding
Section the relations at equal times :
$$
\left[ {\hat b_i,\hat b_j^\dagger } \right]_-=(\;1-
\;\hat \Theta _i\;)\;\delta
_{ij}\;\;\;\;\;\;\;\;\;\;\;\;\;\;\;\;\;\;\;\;\;\;\,\left[ {\hat
f_{i\sigma },\hat f_{j\sigma }^\dagger }
\right]_+=(\;1+\;\hat \Theta _{i\sigma }\;)\;\delta
_{ij} \eqno (76)
$$
and with the same notations as above :
$$
\eqalignno{&\left\langle {\;\left[ {\hat b_q,\hat
b_q^\dagger } \right]_-} \right\rangle=1-
\;\left\langle {\,\hat \Xi _0} \right\rangle &(77)\cr
  &\left\langle {\,\sum\limits_\sigma  {\,\left[ {\hat
f_{(q+k)\sigma },\hat f_{(q+k)\sigma }^\dagger
} \right]_+}} \right\rangle=2+\;\left\langle
{\,\sum\limits_\alpha  {\,\hat \Xi _{0\alpha }}}
\right\rangle=3-\;\left\langle {\,\hat \Xi _0}
\right\rangle &(78)\cr}
$$
Inserting these results in Eqs. (69) and (70), one
finally gets :
$$
\eqalignno{&\sum\limits_\sigma  {\int_{-\infty
}^{+\infty } {{\textstyle{{d\omega } \over {2\pi
}}}\,A_{e\sigma }(k,\omega )}}=1+2\,\delta
-\;\left\langle {\,\hat \Xi _0} \right\rangle &(79)\cr
  &n_e(k)=(1-\delta )\;(\;1-\;\left\langle {\,\hat \Xi
_0} \right\rangle\;)\;\;+\;\;{\textstyle{1 \over
N}}\sum\limits_{q,\sigma } {n_{f\sigma
}(q+k)\;n_b(q)}&(80)\cr
  &{\textstyle{1 \over N}}\sum\limits_k
{n_e(k)}=(1-\delta )\;(\;1-\;\left\langle {\,\hat \Xi
_0} \right\rangle\;+\delta \;)&(81)\cr}
$$

Again, one sees clearly the modifications of the
sum rules and of the results of
Ref. [5] coming from the presence of the $\ \hat
\Theta  \ $'s in the correct canonical relations and
that with :
$$
\left\langle {\,\hat \Xi _0} \right\rangle=\delta
\eqno (82)
$$
one has the correct sum rules for the electron
spectral function :
$$
\sum\limits_\sigma  {\int_{-\infty }^{+\infty }
{{\textstyle{{d\omega } \over {2\pi
}}}\,A_{e\sigma }(q,\omega )}}=1+\delta  \eqno
(83)
$$
and for the electron number :
$$
{\textstyle{1 \over N}}\sum\limits_k
{n_e(k)}=(1-\delta )\;\;\;\;\;\;\;\;\;\;[\;\;{\rm
instead\;of\;the}\;\;(1-\delta ^2)\;\;{\rm
found\;in\;Ref.\;[5]}\;\;] \eqno (84)
$$
However this does not guarantee the existence of
an EFS within the decoupling scheme in the
slave-boson approach of the t-J model, and the
arguments of Ref. [5] against an EFS still
apply.

Furthermore the same arguments as above
lead, using the electron operator
anticommutator, to the operator equality $\ \hat
\Theta _i=\hat b_i^\dagger \hat b_i \ $ .

\eject

{\bf IV - Explicit quantization of the slave-particle
approaches of the t-J model}

\medskip

In the preceding Section, it was shown that the
expected sum rules for the electron
spectral function and for the electron number are
indeed found in the slave-particle approaches
of the t-J model due to the fact that $\ \left\langle
{\,\hat \Theta _i} \right\rangle=\left\langle {\,\hat
\Xi _0} \right\rangle=\delta \ $ , the operator $\
\hat \Theta _i \ $ being the new term
which is present in the canonical relation of the
slave particle when one quantizes the correct
(Dirac) brackets compatible with the constraints.

We furthermore proved using the electron operator
anticommutator that $\ \hat \Theta _i \ $ was the
slave particle number operator. We shall show in
this Section that a direct explicit operator
quantization of the slave-particle approaches of the
t-J model effectively confirms this result
without invoking the electron operator.

\medskip

{\bf a) The slave-fermion representation}

\medskip

Let us first take the slave-fermion case, where we
obtain from the results of Section II:
$$
\left[ {\hat f_i,\hat f_j^\dagger }
\right]_+=(\;1+\;\hat \Theta _i\;)\;\delta
_{ij}\;\;\;\;\;\;\;\;\;\;\;\;\;\;\;\;\;\;\;\;\;\;\,\left[ {\hat
b_{i\alpha },\hat b_{j\alpha }^\dagger } \right]_-
=(\;1-\;\hat \Theta _{i\alpha }\;)\;\delta _{ij} \eqno
(85)
$$
where (cf. Eqs. (53) and (54)) the $\ \hat \Theta  \
$'s correspond to the quantizations :
$$
(\;H_if_i+\;H_i^\dagger f_i^\dagger \;)/D_i\;\;\;\to
\;\;\;\hat \Theta _i\;\;\;\;\;\;\;\;\;\;\;\;\;\;\;(\;G_{i\alpha
}b_{i\alpha }-\;G_{i\alpha }^\dagger b_{i\alpha
}^\dagger \;)/D_i\;\;\;\to \;\;\;\hat \Theta _{i\alpha }
\eqno (86)
$$
However, since $\ D_i\equiv \sum\nolimits_\alpha
{(\;G_{i\alpha }b_{i\alpha }-\;G_{i\alpha
}^\dagger b_{i\alpha }^\dagger
\;)}+(\;H_if_i+\;H_i^\dagger f_i^\dagger \;) \ $ ,
the explicit expressions of the $\ \hat \Theta  \ $
operators in terms of the operators of the fermion
and bosons are {\it a priori} not obvious. We shall
now obtain these expressions through the
structure of the Fock space.

\medskip

 From the slave-particle constraint :
$$
\hat f_i^\dagger \hat f_i+\sum\limits_\alpha  {\hat
b_{i\alpha }^\dagger \hat b_{i\alpha }}=1 \eqno
(87)
$$
the operators $\ \hat f_i^\dagger \hat f_i \ $ and $\
\hat b_{i\alpha }^\dagger \hat b_{i\alpha } \ $
must be particle-number operators satisfying :
$$
(\hat f_i^\dagger \hat f_i)^2=\hat f_i^\dagger \hat
f_i\;\;\;\;\;\;\;\;\;\;\;\;{\rm and}\;\;\;\;\;\;\;\;\;\;\;\;(\hat
b_{i\alpha }^\dagger \hat b_{i\alpha })^2=\hat
b_{i\alpha }^\dagger \hat b_{i\alpha } \eqno (88)
$$
Now it follows from Eq. (85) that :
$$
(\hat f_i^\dagger \hat f_i)^2=\hat f_i^\dagger \hat
f_i-\hat f_i^\dagger (\hat f_i^\dagger \hat f_i-\hat
\Theta _i)\hat f_i\;\;\;\;\;\;\;\;\;\;(\hat b_{i\alpha
}^\dagger \hat b_{i\alpha })^2=\hat b_{i\alpha
}^\dagger \hat b_{i\alpha }+\hat b_{i\alpha
}^\dagger (\hat b_{i\alpha }^\dagger \hat
b_{i\alpha }-\hat \Theta _{i\alpha })\hat b_{i\alpha
} \eqno (89)
$$
and a consistent quantization is thus :
$$
\hat \Theta _i=\hat f_i^\dagger \hat
f_i\;\;\;\;\;\;\;\;\;\;\;\;\;\;\;\hat \Theta _{i\alpha }=\hat
b_{i\alpha }^\dagger \hat b_{i\alpha } \eqno (90)
$$
which from Eq. (85) leads to the relations :
$$
\hat f_i\hat f_i^\dagger
=1\;\;\;\;\;\;\;\;\;\;\;\;\;\;\;\;\;\hat b_{i\alpha }\hat
b_{i\alpha }^\dagger =1 \eqno (91)
$$
We will show below that the dimension of the
Fock space at site {\it i}  is in fact infinite (but
without contradiction with the non double
electron-occupancy at site {\it i} ). Thus Eqs. (91)
are consistent with the fact that the particle-number
operators are not unity. Let us remark that we
have consistently, in accordance with Eqs. (55)
and (87), :
$$
\hat \Theta _i+\sum\limits_\alpha  {\hat \Theta
_{i\alpha }}=\hat f_i^\dagger \hat
f_i+\sum\limits_\alpha  {\hat b_{i\alpha }^\dagger
\hat b_{i\alpha }}=1 \eqno (92)
$$

On the other hand, the fundamental Dirac brackets
compatible with the constraints given
by Eqs. (23-32) lead to the following expressions
:
$$
\eqalignno{&i\,\left\{ {b_{i\alpha }^\dagger
,f_i^\dagger f_i} \right\}_*=b_{i\alpha }^\dagger
\,\Theta _i &(93)\cr
  &i\,\left\{ {f_i^\dagger ,b_{i\alpha }^\dagger
b_{i\alpha }} \right\}_*=f_i^\dagger \,\Theta
_{i\alpha }&(94)\cr
  &i\,\left\{ {b_{i\alpha }^\dagger ,b_{i\beta
}^\dagger b_{i\beta }} \right\}_*=-\;b_{i\beta
}^\dagger \,\delta _{\alpha \beta }+b_{i\alpha
}^\dagger \,\Theta _{i\beta }&(95)\cr}
$$
Quantizing these expressions with the same
ordering of operators gives via Eqs. (90) and (91)
:
$$
\eqalignno{&\hat f_i^\dagger \hat f_i\,\hat
b_{i\alpha }^\dagger =\hat b_{i\alpha }^\dagger
\,(\hat f_i^\dagger \hat f_i-\hat \Theta
_i)=0\;\;\;\;\;\;\;\;\;\;\;\;\;\;\;\;\;\;\;\;{\rm
and\;thus}\;\;\;\;\;\hat f_i\,\hat b_{i\alpha }^\dagger
=0 &(96)\cr
  &\hat b_{i\alpha }^\dagger \hat b_{i\alpha }\hat
f_i^\dagger =\hat f_i^\dagger \,(\hat b_{i\alpha
}^\dagger \hat b_{i\alpha }-\hat \Theta _{i\alpha
})=0\;\;\;\;\;\;\;\;\;\;\;\,\;\;\;\;{\rm
and\;thus}\;\;\;\;\;\hat b_{i\alpha }\hat f_i^\dagger
=0 &(97)\cr
  &\hat b_{i\beta }^\dagger \hat b_{i\beta }\hat
b_{i\alpha }^\dagger =\hat b_{i\alpha }^\dagger
\,(\hat b_{i\beta }^\dagger \hat b_{i\beta }-\hat
\Theta _{i\beta })+\hat b_{i\beta }^\dagger \,\delta
_{\alpha \beta }\;\;\;\;\;\,\,{\rm and\;thus}\;\;\;\;\;\hat
b_{i\beta }\hat b_{i\alpha }^\dagger =\delta
_{\alpha \beta }&(98)\cr}
$$
Using all our results, one easily verifies that the
operators $\ \hat f_i^\dagger \hat f_i \ $ and $\ \hat
b_{i\alpha }^\dagger \hat b_{i\alpha } \ $ are
indeed particle-number operators : for example $\
\hat f_i^\dagger \hat f_i \ $ acting on the fermion
state $\ \hat f_i^\dagger \left| 0 \right\rangle \ $ at
site  {\it i}  has eigenvalue  1  and acting on the
two boson states $\ \hat b_{i\alpha }^\dagger \left|
0 \right\rangle \ $ at site  {\it i}  has eigenvalue  0 .

One can check the consistency of our quantization
: for example quantizing the expression
$$
i\,\left\{ {b_{i\alpha }^\dagger ,b_{i\beta
}b_{i\beta }^\dagger } \right\}_*=-\;b_{i\beta
}^\dagger \,\delta _{\alpha \beta }+\Theta _{i\beta
}\,b_{i\alpha }^\dagger  \eqno (99)
$$
with the same ordering of operators gives :
$$
\eqalignno{&\left[ {\hat b_{i\alpha }^\dagger ,\hat
b_{i\beta }\hat b_{i\beta }^\dagger } \right]_-
=\left[ {\hat b_{i\alpha }^\dagger ,1} \right]_-=0
&(100)\cr
  &\;\;\;\;\;\;\;\;\;\;\;\;\;\;\;\;\;\;\;\;\;\;=-\;\hat b_{i\beta
}^\dagger \,\delta _{\alpha \beta }+\hat b_{i\beta
}^\dagger \hat b_{i\beta }\hat b_{i\alpha
}^\dagger =-\;\hat b_{i\beta }^\dagger \,\delta
_{\alpha \beta }+\hat b_{i\beta }^\dagger \,\delta
_{\alpha \beta }=0 &(101)\cr}
$$

Through an explicit operator quantization of the
slave-fermion approach of the t-J model
we have thus obtained, at site {\it i} , :
$$
\hat f_i\,\hat f_i^\dagger =1\;\;\;\;\;\;\;\;\;\;\;\;\;\;\hat
b_{i\alpha }\hat b_{i\beta }^\dagger =\delta
_{\alpha \beta }\;\;\;\;\;\;\;\;\;\;\;\;\;\;\hat f_i\,\hat
b_{i\alpha }^\dagger =0\;\;\;\;\;\;\;\;\;\;\;\;\;\;\hat
b_{i\alpha }\hat f_i^\dagger =0 \eqno (102)
$$
or in other words :
$$
\eqalignno{&\left[ {\hat f_i,\hat f_j^\dagger }
\right]_+=(\;1+\hat f_i^\dagger \hat f_i\;)\;\delta
_{ij}\;\;\;\;\;\;\;\;\;\;\;\;\;\;\;\left[ {\hat b_{i\alpha
},\hat b_{j\beta }^\dagger } \right]_-=(\;\delta
_{\alpha \beta }-\;\hat b_{i\beta }^\dagger \hat
b_{i\alpha }\;)\;\delta _{ij}&(103)\cr
  &\left[ {\hat f_i,\hat b_{j\alpha }^\dagger }
\right]_-=-\;\hat b_{i\alpha }^\dagger \hat
f_i\;\delta _{ij}\;\;\;\;\;\;\;\;\;\;\;\;\;\;\;\;\;\;\;\;\left[ {\hat
b_{i\alpha },\hat f_j^\dagger } \right]_-=-\;\hat
f_i^\dagger \hat b_{i\alpha }\;\delta
_{ij}&(104)\cr}
$$
Looking in the same way as above at the
quantization of the classical canonical relations
involving only creators or annihilators, we found
either no informations or identities. Such
relations might thus be only identities at the
quantum level. The results of Eq. (102) will
however be sufficient for the purposes of this
paper.

In fact, as a consequence of this study we directly
recover the expressions :
$$
\hat \Theta _i=\hat f_i^\dagger \hat
f_i\;\;\;\;\;\;\;\;\;\;\;\;\;\;\;\left\langle {\,\hat \Theta _i}
\right\rangle=\left\langle {\,\hat \Xi _0}
\right\rangle=\delta  \eqno (105)
$$
which through the analysis of Section III leads to
the conclusion that the expected sum rules for
the electron spectral function and for the electron
number are indeed found in the slave-fermion
approach of the t-J model within the decoupling
approximation. Using Eq. (50) and our result  $\
\hat f_i\hat f_i^\dagger =1 \ $ , one can also see
directly that :
$$
n_e={\textstyle{1 \over N}}\sum\limits_i
{(\sum\limits_\alpha  {\left\langle {\,\hat
b_{i\alpha }^\dagger \hat b_{i\alpha }}
\right\rangle\,})\;\left\langle {\,\hat f_i\hat
f_i^\dagger } \right\rangle}=(1-\;\delta ) \eqno
(106)
$$

It is also important, with the expression of the
electron operator :
$$
\hat c_{i\alpha }^\dagger =\hat b_{i\alpha
}^\dagger \,\hat f_i \eqno (107)
$$
and using our results for the quantization, to
found the exact electron number operator :
$$
\hat n_i=\sum\limits_\alpha  {\hat c_{i\alpha
}^\dagger \hat c_{i\alpha
}^{}}=\sum\limits_\alpha  {\hat b_{i\alpha
}^\dagger \hat f_i\hat f_i^\dagger \hat b_{i\alpha
}}=\sum\limits_\alpha  {\hat b_{i\alpha }^\dagger
\hat b_{i\alpha }}=1-\hat f_i^\dagger \hat f_i
\eqno (108)
$$
and to verify that :
$$
\eqalignno{&\left[ {\hat c_{i\alpha }^\dagger ,\hat
c_{i\beta }^{}} \right]_+=\hat b_{i\alpha
}^\dagger \hat f_i\hat f_i^\dagger \hat b_{i\beta
}+\hat f_i^\dagger \hat b_{i\beta }\hat b_{i\alpha
}^\dagger \hat f_i=\hat b_{i\alpha }^\dagger \hat
b_{i\beta }+\hat f_i^\dagger \hat f_i\,\delta
_{\alpha \beta }&(109)\cr
  &\sum\limits_\alpha  {\left[ {\hat c_{i\alpha
}^\dagger ,\hat c_{i\alpha }^{}}
\right]_+}=1+\hat f_i^\dagger \hat f_i &(110)\cr}
$$
effectively corresponding to the quantization of
Eqs. (37) and (38).

We can now examine, as announced above, the
structure of the Fock space at site {\it i}.
Apart from the three states $\ \left| {f_i}
\right\rangle=\hat f_i^\dagger \left| 0 \right\rangle \
$ and $\ \left| {b_{i\alpha }} \right\rangle=\hat
b_{i\alpha }^\dagger \left| 0 \right\rangle \ $ , we
have an infinite number of states of the form either
$\ \left\| {\left. {f_i} \right\rangle} \right.\equiv
\hat f_i^\dagger (\hat A_i^\dagger )^n\left| 0
\right\rangle \ $ or $\ \left\| {\left. {b_{i\alpha }}
\right\rangle} \right.\equiv \hat b_{i\alpha
}^\dagger (\hat A_i^\dagger )^n\left| 0
\right\rangle \ $ , where
$\ (\hat A_i^\dagger )^n \ $ are products of $\ \hat
b_{i\beta }^\dagger  \ $ and $\ \hat f_i^\dagger  \ $
. The constraint of Eq. (87), counting only the last
created particle, is satisfied for all these states. The
particle content of these states can be found using
a basis of new commuting operators; for example
one has $\ (\hat b_{i\alpha }^\dagger \hat
b_{i\beta }^\dagger \hat b_{i\beta }^{}\hat
b_{i\alpha }^{})\,\hat b_{i\alpha }^\dagger \hat
b_{i\beta }^\dagger \left| 0 \right\rangle=\hat
b_{i\alpha }^\dagger \hat b_{i\beta }^\dagger \left|
0 \right\rangle \ $ . Nevertheless, the states $\ \left|
\;\; \right\rangle \ $ and $\ \left\| {\left. {\;\; }
\right\rangle} \right. \ $ share the same properties
concerning the electron operators :
$$
\eqalign{&\hat c_{i\alpha }^\dagger \left| {f_i}
\right\rangle=\left| {b_{i\alpha }}
\right\rangle\;\;\;\;\;\;\;\;\;\;\;\hat c_{i\alpha
}^\dagger \left| {b_{i\beta }}
\right\rangle=0\;\;\;\;\;\;\;\;\;\;\;\hat c_{i\alpha
}^\dagger \hat c_{i\alpha }^{}\left| {f_i}
\right\rangle=0\;\;\;\;\;\;\;\;\;\;\;\hat c_{i\alpha
}^\dagger \hat c_{i\alpha }^{}\left| {b_{i\beta }}
\right\rangle=\delta _{\alpha \beta }\left| {b_{i\beta
}} \right\rangle\cr
  &\hat c_{i\alpha }^\dagger \left\| {\left. {f_i}
\right\rangle} \right.=\left\| {\left. {b_{i\alpha }}
\right\rangle} \right.\;\;\;\;\;\;\;\,\hat c_{i\alpha
}^\dagger \left\| {\left. {b_{i\beta }} \right\rangle}
\right.=0\;\;\;\;\;\;\;\;\,\hat c_{i\alpha }^\dagger \hat
c_{i\alpha }^{}\left\| {\left. {f_i} \right\rangle}
\right.=0\;\;\;\;\;\;\;\;\;\,\hat c_{i\alpha }^\dagger
\hat c_{i\alpha }^{}\left\| {\left. {b_{i\beta }}
\right\rangle} \right.=\delta _{\alpha \beta }\left\|
{\left. {b_{i\beta }} \right\rangle} \right.\cr}
$$
Thus, there is always either no electron or one
electron at site  {\it i} , as it is also expressed by
the relation $\ \hat c_{i\alpha }^\dagger \hat
c_{i\beta }^\dagger =\hat b_{i\alpha }^\dagger
\,\hat f_i\,\hat b_{i\beta }^\dagger \,\hat f_i=0 \ $
using Eq. (102) . In spite of the infinite dimension
of the Fock space at site  {\it i} , the non double
electron-occupancy at site  {\it i}  is well satisfied,
and furthermore it is equivalent for physical
purposes to use only the sector of the three states
$\ \left| {f_i} \right\rangle \ $ and $\ \left|
{b_{i\alpha }} \right\rangle \ $ .

Let us finally insist on the fact that we have now a
systematic direct algebraic procedure to
find the expression in the slave-particle approach
of any operator initially written in terms of the
electron operators. For instance, since using our
result $\ \hat f_i\hat f_i^\dagger =1 \ $ one has :
$$
\hat c_{i\alpha }^\dagger \hat c_{i\beta }^{}=\hat
b_{i\alpha }^\dagger \hat f_i\hat f_i^\dagger \hat
b_{i\beta }=\hat b_{i\alpha }^\dagger \hat
b_{i\beta } \eqno (111)
$$
we obtain directly the t-J model Hamiltonian of
Eq. (2) in the slave-fermion representation:
$$
\hat H=t\sum\limits_{<ij>,\alpha } {(\hat
b_{i\alpha }^\dagger \hat f_i\hat f_j^\dagger \hat
b_{j\alpha }+h.c.)}+\;J\;\sum\limits_{<ij>}
{{\textstyle{1 \over 2}}(\sum\limits_{\alpha ,\beta
} {\hat b_{i\alpha }^\dagger \hat b_{i\beta }\hat
b_{j\beta }^\dagger \hat b_{j\alpha }}-\;\hat
n_i\,\hat n_j)}+\mu \sum\limits_i {\hat f_i^\dagger
\hat f_i}-\;\mu \,N \eqno (112)
$$
with $\ \hat n_i \ $ given by Eq. (108) , and where
we have added the $\ \mu \ $ chemical potential
term, {\it N}  being the number of lattice sites.

\medskip

{\bf b) The slave-boson representation}

\medskip

The explicit quantization in the slave-boson
representation of the t-J model proceeds
exactly in the same way, and we shall only give
the following results expressed, at site {\it i} , by :
$$
\hat b_i\,\hat b_i^\dagger =1\;\;\;\;\;\;\;\;\;\;\;\;\;\;\hat
f_{i\sigma }\hat f_{i\tau }^\dagger =\delta
_{\sigma \tau }\;\;\;\;\;\;\;\;\;\;\;\;\;\;\hat b_i\,\hat
f_{i\sigma }^\dagger =0\;\;\;\;\;\;\;\;\;\;\;\;\;\;\hat
f_{i\sigma }\hat b_i^\dagger =0 \eqno (113)
$$
or in other terms :
$$
\eqalignno{&\left[ {\hat b_i,\hat b_j^\dagger }
\right]_-=(\;1-\;\hat b_i^\dagger \hat b_i\;)\;\delta
_{ij}\;\;\;\;\;\;\;\;\;\;\;\;\;\;\;\left[ {\hat f_{i\sigma
},\hat f_{j\tau }^\dagger } \right]_+=(\;\delta
_{\sigma \tau }+\;\hat f_{i\tau }^\dagger \hat
f_{i\sigma }\;)\;\delta _{ij}&(114)\cr
  &\left[ {\hat b_i,\hat f_{j\sigma }^\dagger }
\right]_-=-\;\hat f_{i\sigma }^\dagger \hat
b_i\;\delta _{ij}\;\;\;\;\;\;\;\;\;\;\;\;\;\;\;\;\;\;\;\;\left[
{\hat f_{i\sigma },\hat b_j^\dagger } \right]_-=-
\;\hat b_i^\dagger \hat f_{i\sigma }\;\delta
_{ij}&(115)\cr}
$$

We therefore directly recover in this case the
expressions :
$$
\hat \Theta _i=\hat b_i^\dagger \hat
b_i\;\;\;\;\;\;\;\;\;\;\;\;\;\;\;\left\langle {\,\hat \Theta _i}
\right\rangle=\left\langle {\,\hat \Xi _0}
\right\rangle=\delta  \eqno (116)
$$
which through the analysis of Section III leads to
the conclusion that the expected sum rules for
the electron spectral function and for the electron
number are indeed found also in the slave-
boson approach of the t-J model within the
decoupling approximation.

With the expression of the electron operator :
$$
\hat c_{i\sigma }^\dagger =\hat f_{i\sigma
}^\dagger \,\hat b_i \eqno (117)
$$
and using our results for the quantization, let us
also found the exact electron number operator :
$$
\hat n_i=\sum\limits_\sigma  {\hat c_{i\sigma
}^\dagger \hat c_{i\sigma
}^{}}=\sum\limits_\sigma  {\hat f_{i\sigma
}^\dagger \hat b_i\hat b_i^\dagger \hat f_{i\sigma
}}=\sum\limits_\sigma  {\hat f_{i\sigma
}^\dagger \hat f_{i\sigma }}=1-\hat b_i^\dagger
\hat b_i \eqno (118)
$$
to be compared with the ambiguous result using
the (incorrect) naive quantization : since boson
and fermion operators would commute in this
naive quantization, $\ \sum\nolimits_\sigma  {\hat
c_{i\sigma }^\dagger \hat c_{i\sigma }^{}} \ $
could either be written $\ \sum\nolimits_\sigma
{\hat f_{i\sigma }^\dagger \hat f_{i\sigma }\hat
b_i\hat b_i^\dagger } \ $ which using the slave-
particle constraint and the naive commutator of the
boson would give $\ (\;1-\;(\hat b_i^\dagger \hat
b_i)^2\,) \ $ , or be written $\
\sum\nolimits_\sigma  {\hat b_i\hat f_{i\sigma
}^\dagger \hat f_{i\sigma }\hat b_i^\dagger } \ $
which in the same way would give $\ -\;(\hat
b_i^\dagger \hat b_i)\;(\;1+\;(\hat b_i^\dagger \hat
b_i)\,) \ $ .

Finally, since using our result $\ \hat b_i\,\hat
b_i^\dagger =1 \ $ one has :
$$
\hat c_{i\sigma }^\dagger \hat c_{i\tau }^{}=\hat
f_{i\sigma }^\dagger \hat b_i\hat b_i^\dagger \hat
f_{i\tau }=\hat f_{i\sigma }^\dagger \hat f_{i\tau
} \eqno (119)
$$
we obtain directly the t-J model Hamiltonian of
Eq. (2) in the slave-boson representation:
$$
\hat H=t\sum\limits_{<ij>,\sigma } {(\hat
f_{i\sigma }^\dagger \hat b_i\hat b_j^\dagger \hat
f_{j\sigma }+h.c.)}+\;J\;\sum\limits_{<ij>}
{{\textstyle{1 \over 2}}(\sum\limits_{\sigma ,\tau
} {\hat f_{i\sigma }^\dagger \hat f_{i\tau }\hat
f_{j\tau }^\dagger \hat f_{j\sigma }}-\;\hat
n_i\,\hat n_j)}+\mu \sum\limits_i {\hat
b_i^\dagger \hat b_i}-\;\mu \,N \eqno (120)
$$
with $\ \hat n_i \ $ given by Eq. (118) , and where
we have added the $\ \mu \ $ chemical potential
term, {\it N}  being the number of lattice sites.

\medskip

\medskip

{\bf V - Generalization to a slave-boson approach
of the Hubbard model}

\medskip

\medskip

A slave-boson approach was introduced in Ref.
[6] for the Hubbard model, i.e. without
neglecting the doubly occupied sites, and it was
also claimed in Ref. [5] that the sum rule of the
electron number was still violated there. Let us
show explicitly that it is not the case if one uses
our present quantization.

\medskip

In Ref. [6] the electron operator was written in the
finite-{\it U} Hubbard model as :
$$
\hat c_{i\sigma }^\dagger =\hat f_{i\sigma
}^\dagger \,\hat b_{1\,i}^{}+\sigma \,\hat
b_{2\,i}^\dagger \,\hat f_{i\,(-\sigma
)}\;\;\;\;\;\;\;\;\;\;\;\;(\sigma =\pm 1) \eqno (121)
$$
the bosons $\ \hat b_1^\dagger  \ $ and $\ \hat
b_2^\dagger  \ $ describing respectively the empty
and doubly-occupied states. The slave particle
constraint here is :
$$
\hat b_{1\,i}^\dagger \hat b_{1\,i}+\hat
b_{2\,i}^\dagger \hat
b_{2\,i}+\sum\limits_\sigma  {\hat f_{i\sigma
}^\dagger \hat f_{i\sigma }}=1 \eqno (122)
$$

The explicit quantization of this slave-boson
representation of the Hubbard model
proceeds in the same way as shown above for the
slave-particle representation of the t-J model,
and we obtain the results at site {\it i} :
$$
\hat b_{\alpha i}\,\hat b_{\beta i}^\dagger =\delta
_{\alpha \beta }\;\;\;\;\;\;\;\;\;\;\;\;\hat f_{i\sigma
}\hat f_{i\tau }^\dagger =\delta _{\sigma \tau
}\;\;\;\;\;\;\;\;\;\;\;\;\hat b_{\alpha i}\,\hat f_{i\sigma
}^\dagger =0\;\;\;\;\;\;\;\;\;\;\;\;\hat f_{i\sigma }\hat
b_{\alpha i}^\dagger =0 \eqno (123)
$$
or in other terms :
$$
\eqalignno{&\left[ {\hat b_{\alpha i},\hat b_{\beta
j}^\dagger } \right]_-=(\;\delta _{\alpha \beta }-
\;\hat b_{\beta i}^\dagger \hat b_{\alpha
i}\;)\;\delta _{ij}\;\;\;\;\;\;\;\;\;\;\;\;\,\;\left[ {\hat
f_{i\sigma },\hat f_{j\tau }^\dagger }
\right]_+=(\;\delta _{\sigma \tau }+\;\hat f_{i\tau
}^\dagger \hat f_{i\sigma }\;)\;\delta
_{ij}&(124)\cr
  &\left[ {\hat b_{\alpha i},\hat f_{j\sigma
}^\dagger } \right]_-=-\;\hat f_{i\sigma }^\dagger
\hat b_{\alpha i}\;\delta
_{ij}\;\;\;\;\;\;\;\;\;\;\;\;\;\;\;\;\;\;\;\;\;\;\;\;\;\left[ {\hat
f_{i\sigma },\hat b_{\alpha j}^\dagger } \right]_-
=-\;\hat b_{\alpha i}^\dagger \hat f_{i\sigma
}\;\delta _{ij}&(125)\cr}
$$
where $\ \alpha  \ $ or $\ \beta = 1 , 2  \ $ and  $\
\sigma \ $ or $\ \tau = \pm 1 \ $ .

The electron field being gauge invariant, one
verifies as above that at the classical level
one can use the initial brackets of the {\it b} 's and
{\it f} 's to compute the Dirac brackets of the
electron field :
$$
i\;\left\{ {c_{i\sigma }^\dagger ,c_{i\tau }^{}}
\right\}=i\;\left\{ {c_{i\sigma }^\dagger ,c_{i\tau
}^{}} \right\}_*=\;\delta _{\sigma \tau } \eqno
(126)
$$
while at the quantum level our results give :
$$
\eqalignno{&\hat c_{i\sigma }^\dagger \hat
c_{i\tau }^{}=\hat f_{i\sigma }^\dagger \hat
f_{i\tau }+\hat b_{2\,i}^\dagger \hat
b_{2\,i}^{}\,\delta _{\sigma \tau }&(127)\cr
  &\hat c_{i\tau }^{}\hat c_{i\sigma }^\dagger
=\sigma \tau \,\hat f_{i(-\tau )}^\dagger \hat f_{i(-
\sigma )}+\hat b_{1\,i}^\dagger \hat
b_{1\,i}^{}\,\delta _{\sigma \tau }&(128)\cr}
$$
which effectively leads to :
$$
\left[ {\hat c_{i\tau }^{},\hat c_{i\sigma
}^\dagger } \right]_+=\;\delta _{\sigma \tau }
\eqno (129)
$$

One has from Eq. (127) the expression of the
exact electron number operator :
$$
\sum\limits_\sigma  {\hat c_{i\sigma }^\dagger
\hat c_{i\sigma }^{}}=1-\;(\hat b_{1\,i}^\dagger
\hat b_{1\,i}^{}-\hat b_{2\,i}^\dagger \hat
b_{2\,i}^{}) \eqno (130)
$$
Let us define $\ \Delta  \ $ as the average number
of empty sites,  $\ d \ $  the average number of
doubly occupied sites  and $\ \delta  \ $ the hole
doping concentration :
$$
\Delta =\left\langle {\hat b_{1\,i}^\dagger \hat
b_{1\,i}^{}}
\right\rangle\;\;\;\;\;\;\;\;\;\;\;\;\;\;\;d=\left\langle {\hat
b_{2\,i}^\dagger \hat b_{2\,i}^{}}
\right\rangle\;\;\;\;\;\;\;\;\;\;\;\;\;\;\;\delta =\Delta -\;d
\eqno (131)
$$
The total electron number per site is then correctly
obtained from Eq. (130) as :
$$n_e=(1-\;\delta ) \eqno (132)$$

Let us again emphasize that the expression of the
Hamiltonian in the slave-particle
approach is obtained, using our results, through a
direct algebraic procedure. Eqs. (127) and
(123) lead to :
$$
\hat n_{i+}\hat n_{i-}=(\hat f_{i+}^\dagger \hat
f_{i+}+\hat b_{2\,i}^\dagger \hat
b_{2\,i}^{})\,(\hat f_{i-}^\dagger \hat f_{i-}+\hat
b_{2\,i}^\dagger \hat b_{2\,i}^{})=\hat
b_{2\,i}^\dagger \hat b_{2\,i}^{} \eqno (133)
$$
and we obtain directly the Hubbard Hamiltonian
of Eq. (1) with nearest-neighbor hopping in
the slave-boson representation of Eq. (121) :
$$
\eqalignno{&\hat H=t\sum\limits_{<ij>,\sigma }
{\hat f_{i\sigma }^\dagger (\hat b_{1\,i}^{}\hat
b_{1\,j}^\dagger -\hat b_{2\,i}^{}\hat
b_{2\,j}^\dagger )\hat f_{j\sigma
}}+t\;\sum\limits_{<ij>} {[\,\hat f_{i+}^\dagger
\hat f_{j-}^\dagger (\hat b_{1\,i}^{}\hat
b_{2\,j}^{}+\hat b_{1\,j}^{}\hat
b_{2\,i}^{})+h.c.\,]}\cr
  &\;\;\;\;\;\,+\;U\sum\limits_i {\hat
b_{2\,i}^\dagger \hat b_{2\,i}^{}}+\mu
\sum\limits_i {(\hat b_{1\,i}^\dagger \hat
b_{1\,i}^{}-\hat b_{2\,i}^\dagger \hat
b_{2\,i}^{})}-\;\mu \,N &(134)\cr}
$$
where we have added the $\ \mu \ $ chemical
potential term,  {\it N}  being the number of lattice
sites.

The decoupling approximation for the Matsubara
electron Green's function in imaginary
time is here expressed by :
$$
E_\sigma (r,\tau )=-\;F_\sigma (r,\tau )\;B_1(-r,-
\tau )+F_{-\sigma }(-r,-\tau )\;B_2(r,\tau ) \eqno
(135)
$$

On one hand, one obtains the sum rule for the
electron spectral function within the
decoupling approximation :
$$
\eqalignno{&\sum\limits_\sigma  {\int_{-\infty
}^{+\infty } {{\textstyle{{d\omega } \over {2\pi
}}}\,A_{e\sigma }(k,\omega )}}&(136)\cr
  &\;\;\;\;\;\;\;\;={\textstyle{1 \over
N}}\sum\limits_q {n_{b_1}(q)\;\left\langle
{\,\sum\limits_\sigma  {\,\left[ {\hat
f_{(q+k)\sigma },\hat f_{(q+k)\sigma }^\dagger
} \right]_+}} \right\rangle}+{\textstyle{1 \over
N}}\sum\limits_{q,\sigma } {n_{f\sigma
}(q+k)\;\left\langle {\;\left[ {\hat b_{1\,q},\hat
b_{1\,q}^\dagger } \right]_-} \right\rangle}\cr
  &\;\;\;\;\;\;\;\;+{\textstyle{1 \over
N}}\sum\limits_{q,\sigma } {n_{f\sigma
}(q)\;\left\langle {\;\left[ {\hat b_{2\,(q+k)},\hat
b_{2\,(q+k)}^\dagger } \right]_-}
\right\rangle}+{\textstyle{1 \over
N}}\sum\limits_q {n_{b_2}(q+k)\;\left\langle
{\,\sum\limits_\sigma  {\,\left[ {\hat f_{q\sigma
},\hat f_{q\sigma }^\dagger } \right]_+}}
\right\rangle}\cr}
$$
which using our quantization gives :
$$
\eqalignno{&\sum\limits_\sigma  {\int_{-\infty
}^{+\infty } {{\textstyle{{d\omega } \over {2\pi
}}}\,A_{e\sigma }(k,\omega )}}&(137)\cr
  &\;\;\;\;\;\;\;\;=\Delta \,(3-\Delta -d)+(1-\Delta
-d)\,(1-d)+(1-\Delta -d)\,(1-\Delta )+d\,(3-\Delta
-d)=2\cr}
$$
as it should be from Eq. (129) .

On the other hand, Eq. (135)  gives the expression
for the number of electrons in state  {\it k}
within the decoupling approximation :
$$
\eqalignno{&n_e(k)\;\;={\textstyle{1 \over
N}}\sum\limits_{q,\sigma } {\left[ {\left\langle
{\hat f_{(q+k)\sigma }^\dagger \hat
f_{(q+k)\sigma }^{}} \right\rangle\left\langle
{\hat b_{1\,q}\hat b_{1\,q}^\dagger }
\right\rangle+\left\langle {\hat
b_{2\,(q+k)}^\dagger \hat b_{2\,(q+k)}^{}}
\right\rangle\left\langle {\hat f_{q\sigma }\hat
f_{q\sigma }^\dagger } \right\rangle} \right]}\cr
  &\;\;\;\;\;\;\;\;\,={\textstyle{1 \over
N}}\sum\limits_{q,\sigma } {n_{f\sigma
}(q+k)\;\left\langle {\;\left[ {\hat b_{1\,q},\hat
b_{1\,q}^\dagger } \right]_-}
\right\rangle}\;\;+\;\;{\textstyle{1 \over
N}}\sum\limits_{q,\sigma } {n_{f\sigma
}(q+k)\;n_{b_1}(q)}&(138)\cr
  &\;\;\;\;\;\;\;\;\,+{\textstyle{1 \over
N}}\sum\limits_{q,\sigma }
{n_{b_2}(q+k)\;\left\langle {\;\left[ {\hat
f_{q\sigma },\hat f_{q\sigma }^\dagger }
\right]_+} \right\rangle}\;\;-\;\;{\textstyle{1 \over
N}}\sum\limits_{q,\sigma }
{n_{b_2}(q+k)\;n_{f\sigma }(q)}\cr}
$$
Using our results for the quantization of the $\ b_1
\ $ boson and the $\ f \ $ fermion, one obtains :
$$
n_e(k)=(1-\Delta )^2-d^2+2d\;\;+\;\;{\textstyle{1
\over N}}\sum\limits_{q,\sigma } {n_{f\sigma
}(q+k)\;n_{b_1}(q)}\;\;-\;\;{\textstyle{1 \over
N}}\sum\limits_{q,\sigma }
{n_{b_2}(q+k)\;n_{f\sigma }(q)} \eqno (139)
$$
 from which we indeed get the expected sum rule
of the electron number :
$$
{\textstyle{1 \over N}}\sum\limits_k
{n_e(k)}=(1-\Delta +d)=(1-\delta ) \eqno (140)
$$
This last result also directly follows from Eq.
(135) :
$$
\eqalignno{&n_e={\textstyle{1 \over
N}}\sum\limits_i {\left\langle {\,\hat b_{1\,i}\hat
b_{1\,i}^\dagger }
\right\rangle\;(\sum\limits_\sigma  {\left\langle
{\,\hat f_{i\sigma }^\dagger \hat f_{i\sigma }}
\right\rangle\,})}+{\textstyle{1 \over
N}}\sum\limits_i {(\sum\limits_\sigma
{\left\langle {\,\hat f_{i\sigma }\hat f_{i\sigma
}^\dagger } \right\rangle\,})\;\left\langle {\,\hat
b_{2\,i}^\dagger \hat b_{2\,i}} \right\rangle}\cr
  &\;\;\;\,=1\;(1-\Delta -d)+2\;d=(1-\delta )
&(141)\cr}
$$
using the results of our quantization, instead of :
$$
(1+\Delta )\;(1-\Delta -d)+(2-(1-\Delta -d))\;d=(1-
\Delta ^2+d^2)
$$
using the (incorrect) naive quantization of Refs.
[5] and [6] .

\medskip

\medskip

{\bf VI - Conclusions}

\medskip

\medskip

In this paper, we have first presented at the
classical level the consistent Hamiltonian
formulation of models having a slave-particle
constraint for their fields. Due to this constraint,
the naive canonical relations are replaced by
modified canonical relations which are compatible
with the constraint. This is achieve through the
use of Dirac brackets, after fixing the gauge
generated by the slave-particle constraint.

We have then shown at the quantum level, for the
slave-fermion and the slave-boson
representations of the t-J model and for a slave-
boson representation of the Hubbard model,
that a consistent quantization of these modified
canonical relations changes the naive sum rules
for the slave particles. These naive sum rules used
in Ref. [5] were there shown to lead to
difficulties in these slave-particle approaches for
the t-J and Hubbard models, coming from the
fact that the sum rule of the electron number was
violated within a decoupling approximation.
On the contrary, we find that, using our
quantization and modified sum rules for the slave
particles, the sum rule of the electron number is in
fact well obeyed. On the other hand, we
obtain a systematic direct algebraic procedure to
find the exact expression in a slave-particle
approach of any operator, e.g. the Hamiltonian,
initially written in terms of the electron
operators.

We thus show that one has to be careful about the
canonical relations when using a direct
quantum operator approach for slave-particle
theories. Of course we are not concerned for these
theories neither with the functional integral
approach nor with the Abrikosov [14] method
used in some slave-particle theories [15].

\medskip

\medskip

\medskip

${\underline {\rm Acknowledgements}}$

We would like to acknowledge useful
conversations with D. Arnaudon, F. Delduc,
L. Frappat and Th. Jolicoeur about this paper.

\eject

\centerline {{\bf REFERENCES}}

\vskip 20mm

[1] P.W. Anderson, Science {\bf 235}, 1196
(1987).

[2] F.C. Zhang and T.M. Rice, Phys. Rev. B {\bf
37}, 3759 (1988).

[3] See, e.g., the review, L. Yu, in {\it Recent
Progress in Many-Body Theories}, edited
by T.L. Ainsworth {\it et al.} (Plenum, New
York, 1992), Vol. 3, p. 157.

[4] M. Ogata and H. Shiba, Phys. Rev. B {\bf
41}, 2326 (1990) ; H. Shiba and M. Ogata,
Int. J. Mod. Phys. B {\bf 5}, 31 (1991).

[5] Shiping Feng, J.B. Wu, Z.B. Su and L. Yu,
Phys. Rev. B {\bf 47}, 15192 (1993).

[6] Z. Zou and P.W. Anderson, Phys. Rev. B
{\bf 37}, 627 (1988).

[7] P.A.M. Dirac, Can. J. Math. {\bf 2}, 129
(1950) ; P.A.M. Dirac, {\it Lectures on Quantum
Mechanics} (Belfer Graduate School of Science,
Yeshiva University,  New York, 1964).

[8] A. Hanson, T. Regge and C. Teitelboim, {\it
Constrained Hamiltonian Systems} (Aca-demia
Nazionale dei Lincei, Rome, 1976).

[9] K. Sundermeyer, {\it Constrained Dynamics}
(Springer Lectures Vol. 169, Springer
Verlag, New York, 1982).

[10] P.B. Wiegmann, Phys. Rev. Lett. {\bf 60},
821 (1988) ; D. Frster, Phys. Rev. Lett.
{\bf 63}, 2140 (1989) ; S. Sarkar, J. Phys. A
{\bf 24}, 1137 (1991).

[11] J. Hubbard, Proc. R. Soc. London Ser. A
{\bf 285}, 542 (1965).

[12] A.A. Abrikosov, L.P. Gorkov and I.E.
Dzyaloshinski, {\it Methods of Quantum
Field Theory in Statistical Physics} (Dover, New
York, 1975).

[13] G.D. Mahan, {\it Many-Particle Physics}
2nd ed. (Plenum, New York, 1990).

[14] A.A Abrikosov, Physics {\bf 2}, 5 (1965).

[15] S.E. Barnes, J. Phys. F {\bf 6}, 1375
(1976) and {\bf 7}, 2637 (1977) ; P. Coleman,
Phys. Rev. B {\bf 28}, 5255 (1983) and {\bf
29}, 3035 (1984).

\vskip 20mm

\eject

\centerline {{\bf Appendix}}

\medskip

\medskip

In this Appendix, we show how the brackets of
Eq. (7) and the first class constraints of
Eq. (8) are obtained and we present the expression
of the corresponding Hamiltonian.

Let us consider the classical Lagrangian (written in
real time) for  {\it n}  bosons $\ b_\alpha \ $
 and  {\it m}  fermions $\ f_\sigma \ $(Grassmann
variables) on a lattice :
$$
\eqalignno{&L=\;\;\;\;\;i{\textstyle{1 \over
2}}(\lambda +1)\sum\limits_{i,\alpha }
{b_{i\alpha }^\dagger \partial _tb_{i\alpha
}}+i{\textstyle{1 \over 2}}(\lambda
-1)\sum\limits_{i,\alpha } {\partial _tb_{i\alpha
}^\dagger b_{i\alpha }}\cr
  &\;\;\;\;\;\;+\;\,i{\textstyle{1 \over 2}}(\lambda
'+1)\sum\limits_{i,\sigma } {f_{i\sigma
}^\dagger \partial _tf_{i\sigma }}+i{\textstyle{1
\over 2}}(\lambda '-1)\sum\limits_{i,\sigma }
{\partial _tf_{i\sigma }^\dagger f_{i\sigma
}}&(A1)\cr
  &\;\;\;\;\;\;+\;X(b^\dagger ,b,f^\dagger
,f)+\sum\limits_i {\Lambda
_i\;(\;\sum\limits_\alpha  {b_{i\alpha }^\dagger
b_{i\alpha }}+\sum\limits_\sigma  {f_{i\sigma
}^\dagger f_{i\sigma }}-1\;)}\cr}
$$
the Lagrangian of Eq. (5) being obtained for $\
\lambda =\lambda '=1 \ $. The canonical momenta
are then :
$$
\eqalignno{&\bar b_{i\alpha }=i{\textstyle{1
\over 2}}(\lambda +1)\,b_{i\alpha }^\dagger
\;\;\;\;\;\;\;\;\;\;\;\;\;\;\bar b_{i\alpha }^\dagger
=i{\textstyle{1 \over 2}}(\lambda -1)\,b_{i\alpha
}\cr
  &\overline f_{i\sigma }=-\,i{\textstyle{1 \over
2}}(\lambda '+1)\,f_{i\sigma }^\dagger
\;\;\;\;\;\;\;\;\;\overline f_{i\sigma }^\dagger
=i{\textstyle{1 \over 2}}(\lambda '-1)\,f_{i\sigma
}\;\;\;\;\;\;\;\;\;\;\Pi _i=0 &(A2)\cr}
$$
The canonical graded (see Eq. (9)) Poisson
brackets, the non zero ones being :
$$
\eqalignno{&\left\{ {b_{i\alpha },\bar b_{j\beta
}} \right\}=\left\{ {b_{i\alpha }^\dagger ,\bar
b_{j\beta }^\dagger } \right\}=\delta _{ij}\,\delta
_{\alpha \beta }\cr
  &\left\{ {f_{i\sigma },\overline f_{j\tau }}
\right\}=\left\{ {f_{i\sigma }^\dagger ,\overline
f_{j\tau }^\dagger } \right\}=-\,\delta _{ij}\,\delta
_{\sigma \tau }\;\;\;\;\;\;\;\;\;\;\;\;\;\;\;\left\{ {\Lambda
_i,\Pi _j} \right\}=\delta _{ij}&(A3)\cr}
$$
are however not compatible with the expressions
of the canonical momenta. Following the
general procedure of Dirac [7], one has to
consider here as Hamiltonian :
$$
\eqalignno{&H_1=\;-\,X(b^\dagger ,b,f^\dagger
,f)-\sum\limits_i {\Lambda
_i\;(\;\sum\limits_\alpha  {b_{i\alpha }^\dagger
b_{i\alpha }}+\sum\limits_\sigma  {f_{i\sigma
}^\dagger f_{i\sigma }}-1\;)}\cr
  &\;\;\;\;\;\;\;\;\;\;+\sum\limits_{i,\alpha }
{(B_{i\alpha }u_{i\alpha }+u_{i\alpha }^\dagger
B_{i\alpha }^\dagger )}+\sum\limits_{i,\sigma }
{(F_{i\sigma }v_{i\sigma }+v_{i\sigma
}^\dagger F_{i\sigma }^\dagger )}+\sum\limits_i
{\Pi _iw_i}&(A4)\cr}
$$
where the $\ u\;,\;u^\dagger ,\;v\;,\;v^\dagger ,\;w
\ $ are unknown (at this stage independent of the
fields and of the canonical momenta) coefficients
and where :
$$
\eqalignno{&B_{i\alpha }\equiv \bar b_{i\alpha }-
\;i{\textstyle{1 \over 2}}(\lambda +1)\,b_{i\alpha
}^\dagger \;\;\;\;\;\;\;\;\;\;\;\;\;B_{i\alpha }^\dagger
\equiv \bar b_{i\alpha }^\dagger -\;i{\textstyle{1
\over 2}}(\lambda -1)\,b_{i\alpha
}\;\;\;\;\;\;\;\;\;\;\;\;\;\Pi _i\cr
  &F_{i\sigma }\equiv \overline f_{i\sigma
}+i{\textstyle{1 \over 2}}(\lambda
'+1)\,f_{i\sigma }^\dagger
\;\;\;\;\;\;\;\;\;\;\;\;\;F_{i\sigma }^\dagger \equiv
\overline f_{i\sigma }^\dagger -\;i{\textstyle{1
\over 2}}(\lambda '-1)\,f_{i\sigma }&(A5)\cr}
$$
are {\it primary constraints} which are weakly
zero, meaning that one has to set the constraints
only after computing all the brackets. In order to
have a consistent system, we require the time
derivatives of these primary constraints to be
weakly zero ($\ \approx 0 \ $), which gives :
$$
\eqalignno{&\left\{ {B_{i\alpha },H_1}
\right\}={{\partial X} \over {\partial b_{i\alpha
}}}+\Lambda _ib_{i\alpha }^\dagger
-\;i\,u_{i\alpha }^\dagger \approx 0\;\;\;\;\;\;\;\left\{
{B_{i\alpha }^\dagger ,H_1} \right\}={{\partial
X} \over {\partial b_{i\alpha }^\dagger
}}+\Lambda _ib_{i\alpha }+i\,u_{i\alpha
}\approx 0\;\;\;\; &(A6)\cr
  &\left\{ {F_{i\sigma },H_1} \right\}={{\partial
X} \over {\partial f_{i\sigma }}}-\;\Lambda
_if_{i\sigma }^\dagger +i\,v_{i\sigma }^\dagger
\approx 0\;\;\;\;\;\;\;\;\left\{ {F_{i\sigma }^\dagger
,H_1} \right\}={{\partial X} \over {\partial
f_{i\sigma }^\dagger }}+\Lambda _if_{i\sigma }-
\;i\,v_{i\sigma }\approx 0\;\;\;\; &(A7)\cr
  &\left\{ {\Pi _i,H_1} \right\}=\sum\limits_\alpha
{b_{i\alpha }^\dagger b_{i\alpha
}}+\sum\limits_\sigma  {f_{i\sigma }^\dagger
f_{i\sigma }}-1\equiv \Phi _i\approx 0 &(A8)\cr}
$$
Eqs. (A6) and (A7) determine the $\
u\;,\;u^\dagger ,\;v\;,\;v^\dagger \ $ and Eq. (A8)
gives the slave-particle constraint which appears
as a {\it secondary constraint}. Repeating the
process, we require the time derivative of this
secondary constraint to be weakly zero, which
gives :
$$
\left\{ {\Phi _i,H_1} \right\}=\sum\limits_\alpha
{(b_{i\alpha }^\dagger u_{i\alpha }+u_{i\alpha
}^\dagger b_{i\alpha })}\;+\sum\limits_\sigma
{(-\;f_{i\sigma }^\dagger v_{i\sigma
}+v_{i\sigma }^\dagger f_{i\sigma })}\approx 0
\eqno (A9)
$$
Using the expressions of the $\ u\;,\;u^\dagger
,\;v\;,\;v^\dagger \ $ given by Eqs. (A6) and (A7),
the Eq. (A9) reads :
$$
\sum\limits_\alpha  {(b_{i\alpha }^\dagger
{{\partial X} \over {\partial b_{i\alpha }^\dagger
}}-\;{{\partial X} \over {\partial b_{i\alpha
}}}b_{i\alpha })}\;+\sum\limits_\sigma
{(f_{i\sigma }^\dagger {{\partial X} \over
{\partial f_{i\sigma }^\dagger }}+{{\partial X}
\over {\partial f_{i\sigma }}}f_{i\sigma })}=0
\eqno (A10)
$$
which is identically satisfied if  {\it X}  is a
function of $\ b^\dagger b \ $ and $\ f^\dagger f \
$.

One recall that a quantity is called {\it first class} if
its bracket with each of the (primary and
secondary) constraints is weakly zero, and {\it
second class} if at least one of these brackets is
not weakly zero. $\ \Pi _i \ $ is then first class,
while our other constraints are {\it a priori}
second class since :
$$
\eqalignno{&\left\{ {B_{i\alpha },B_{j\beta
}^\dagger } \right\}=-\;i\,\delta _{ij}\,\delta
_{\alpha \beta }\;\;\;\;\;\;\;\;\;\;\;\;\;\;\;\;\;\;\left\{
{F_{i\sigma },F_{j\tau }^\dagger } \right\}=-
\;i\,\delta _{ij}\,\delta _{\sigma \tau }&(A11)\cr
  &\left\{ {B_{i\alpha },\Phi _j} \right\}=-
\;b_{i\alpha }^\dagger \delta
_{ij}\;\;\;\;\;\;\;\;\;\;\;\;\;\;\;\;\;\;\;\;\;\;\;\;\,\left\{
{B_{i\alpha }^\dagger ,\Phi _j} \right\}=-
\;b_{i\alpha }^{}\delta _{ij}&(A12)\cr
  &\left\{ {F_{i\sigma },\Phi _j}
\right\}=+\;f_{i\sigma }^\dagger \delta
_{ij}\;\;\;\;\;\;\;\;\;\;\;\;\;\;\;\;\;\;\;\;\;\;\;\;\,\,\left\{
{F_{i\sigma }^\dagger ,\Phi _j} \right\}=-
\;f_{i\sigma }^{}\delta _{ij}&(A13)\cr}
$$
However, the following linear combination of
these second class constraints :
$$
\eqalignno{&\Psi _i\equiv \Phi
_i+i\sum\limits_\alpha  {(-\;B_{i\alpha
}b_{i\alpha }+b_{i\alpha }^\dagger B_{i\alpha
}^\dagger )}+i\sum\limits_\sigma  {(F_{i\sigma
}f_{i\sigma }+f_{i\sigma }^\dagger F_{i\sigma
}^\dagger )} &(A14)\cr
  &\;\;\;\,=\;\;\,i\sum\limits_\alpha  {(-\;\bar
b_{i\alpha }b_{i\alpha }+\bar b_{i\alpha
}^\dagger b_{i\alpha }^\dagger
)}+i\sum\limits_\sigma  {(\overline f_{i\sigma
}f_{i\sigma }-\;\overline f_{i\sigma }^\dagger
f_{i\sigma }^\dagger )}-\;1 &(A15)\cr}
$$
(which is weakly equal to $\ \Phi _i \ $) can be
verified to be first class.

Thus, $\Pi _i$ and $\Psi _i$ are first class; and
$B_{i\alpha }$ , $B_{i\alpha }^\dagger $ ,
$F_{i\sigma }$ and $F_{i\sigma }^\dagger $
(which are such that no linear combination of them
is first class) are second class. Defining $\varphi
_1\equiv B_{i\alpha }$ , $\varphi _2\equiv
B_{i\alpha }^\dagger $ , $\varphi _3\equiv
F_{i\sigma }$ , $\varphi _4\equiv F_{i\sigma
}^\dagger $ , the matrix $C_{ab}\equiv \left\{
{\varphi _a,\varphi _b} \right\}$  is non singular.
Systematic use of the standard Dirac bracket [7] of
two quantities {\it A}  and {\it B} :
$$
\left\{ {A,B} \right\}_D\equiv \left\{ {A,B}
\right\}-\sum\limits_{a,b} {\left\{ {A,\varphi _a}
\right\}\,(C^{-1})_{ab}\,\left\{ {\varphi _b,B}
\right\}} \eqno (A16)
$$
then allows to set all these second class constraints
strongly to zero because the Dirac bracket of
anything with a second class constraint vanishes.

One then finds the following results :
$$
\eqalignno{&\left\{ {\bar b_{i\alpha }^\dagger
,b_{j\beta }^\dagger } \right\}_D={\textstyle{1
\over 2}}(\lambda -1)\,\delta _{ij}\,\delta _{\alpha
\beta }\;\;\;\;\;\;\;\;\;\;\;\;\;\;\;\left\{ {\bar b_{i\alpha
}^{},b_{j\beta }^{}} \right\}_D=-{\textstyle{1
\over 2}}(\lambda +1)\,\delta _{ij}\,\delta
_{\alpha \beta } &(A17)\cr
  &\left\{ {\bar b_{i\alpha }^\dagger ,\bar
b_{j\beta }^{}} \right\}_D=i{\textstyle{1 \over
4}}(\lambda ^2-1)\,\delta _{ij}\,\delta _{\alpha
\beta }\;\;\;\;\;\;\;\;\;\;\;\;\left\{ {b_{i\alpha
},b_{j\beta }^\dagger } \right\}_D=-\;i\;\delta
_{ij}\delta _{\alpha \beta } &(A18)\cr
  &\left\{ {\overline f_{i\sigma }^\dagger ,f_{j\tau
}^\dagger } \right\}_D={\textstyle{1 \over
2}}(\lambda '-1)\,\delta _{ij}\,\delta _{\sigma \tau
}\;\;\;\;\;\;\;\;\;\;\;\;\;\;\left\{ {\overline f_{i\sigma
},f_{j\tau }^{}} \right\}_D=-{\textstyle{1 \over
2}}(\lambda '+1)\,\delta _{ij}\,\delta _{\sigma
\tau } &(A19)\cr
  &\left\{ {\overline f_{i\sigma }^\dagger
,\overline f_{j\tau }} \right\}_D=-\,i{\textstyle{1
\over 4}}(\lambda '^{\,2}-1)\,\delta _{ij}\,\delta
_{\sigma \tau }\;\;\;\;\;\;\;\left\{ {f_{i\sigma
},f_{j\tau }^\dagger } \right\}_D=-\;i\;\delta
_{ij}\delta _{\sigma \tau } &(A20)\cr
  &\left\{ {\Lambda _i,\Pi _j} \right\}_D=\delta
_{ij} &(A21)\cr}
$$
Since we can now set the second class constraints
strongly to zero, i.e. use the expressions of
$\;\bar b_{i\alpha }\,,\;\bar b_{i\alpha }^\dagger
\,,\;\overline f_{i\sigma }\,,\;\overline f_{i\sigma
}^\dagger $  given by Eqs. (A2), a choice of
independent non zero canonical relations is :
$$
\left\{ {b_{i\alpha },b_{j\beta }^\dagger }
\right\}_D=-\;i\;\delta _{ij}\delta _{\alpha \beta
}\;\;\;\;\;\;\;\;\;\left\{ {f_{i\sigma },f_{j\tau
}^\dagger } \right\}_D=-\;i\;\delta _{ij}\delta
_{\sigma \tau }\;\;\;\;\;\;\;\;\;\left\{ {\Lambda _i,\Pi
_j} \right\}_D=\delta _{ij} \eqno (A22)
$$
which are those of Eq. (7) of Section II where we
have drop for convenience the subscript  {\it D} .
Let us note that the parameters $\lambda $ and
$\lambda '$ of the Lagrangian (A1) no longer
appear, and can thus, as usual, be taken as unity
in the Lagrangian.

At this stage, the Hamiltonian is :
$$
H_2=-\,X(b^\dagger ,b,f^\dagger ,f)-
\sum\limits_i {\Lambda _i\;(\;\sum\limits_\alpha
{b_{i\alpha }^\dagger b_{i\alpha
}}+\sum\limits_\sigma  {f_{i\sigma }^\dagger
f_{i\sigma }}-1\;)}+\sum\limits_i {\Pi _iw_i}
\eqno (A23)
$$
and one now has to deal, as shown in Section II,
with the first class constraints :
$$
\Phi _i\equiv \sum\limits_\alpha  {b_{i\alpha
}^\dagger b_{i\alpha }}+\sum\limits_\sigma
{f_{i\sigma }^\dagger f_{i\sigma }}-1\approx
0\;\;\;\;\;\;\;\;\;\;\;\;\;\;\;\;\;\Pi _i\approx 0 \eqno (A24)
$$
Let us note that we have consistently for example
$\left\{ {b_{i\alpha },H_1} \right\}=\left\{
{b_{i\alpha },H_2} \right\}_D$  since :
$$
\eqalignno{&\left\{ {b_{i\alpha },H_1}
\right\}=u_{i\alpha } &(A25)\cr
  &\left\{ {b_{i\alpha },H_2}
\right\}_D=i{{\partial X} \over {\partial
b_{i\alpha }^\dagger }}+i\,\Lambda _ib_{i\alpha
} &(A26)\cr}
$$
which are indeed equal using Eq. (A6) .

\bye